\documentclass{article}

\usepackage{amssymb,amsfonts,amsmath,stmaryrd}
\usepackage{cite,enumerate,float,indentfirst}
\def\be{\begin{eqnarray}}
\def\ee{\end{eqnarray}}
\def\nn{\nonumber}

\def\p{\partial}

\def\Tr{{\rm Tr}\,}

\def\ent{{\rm floor}}
\def\fr{{\rm frac}}

\hoffset= - 1.0in         % switch off for draft style
\begin{document}
\title{\bf
Algebra of quantum ${\cal C}$-polynomials
}

\author{
{\bf Andrei Mironov$^{a,b,c}$}\footnote{mironov@lpi.ru; mironov@itep.ru},
\ and  \  {\bf Alexei Morozov$^{d,b,c}$}\thanks{morozov@itep.ru}
}
\date{ }

\maketitle

\vspace{-5.0cm}

\begin{center}
\hfill FIAN/TD-13/20\\
\hfill IITP/TH-13/20\\
\hfill ITEP/TH-18/20\\
\hfill MIPT/TH-12/20
\end{center}

\vspace{2.cm}

\begin{center}
$^a$ {\small {\it Lebedev Physics Institute, Moscow 119991, Russia}}\\
$^b$ {\small {\it ITEP, Moscow 117218, Russia}}\\
$^c$ {\small {\it Institute for Information Transmission Problems, Moscow 127994, Russia}}\\
$^d$ {\small {\it MIPT, Dolgoprudny, 141701, Russia}}
\end{center}

\vspace{.0cm}

\begin{abstract}
Knot polynomials colored with symmetric representations of $SL_q(N)$ satisfy difference equations as functions of representation
parameter, which look like quantization of classical ${\cal A}$-polynomials.
However, they are quite difficult to derive and investigate.
Much simpler should be the equations for coefficients of differential expansion
nicknamed quantum ${\cal C}$-polynomials.
It turns out that, for each knot, one can actually derive two difference equations of a finite order for these coefficients,
those with shifts in spin $n$ of the representation and in $A=q^N$.
Thus, the ${\cal C}$-polynomials are much richer and form an entire ring.
We demonstrate this with the examples of various defect zero knots,
mostly discussing the entire twist family.
\end{abstract}

\section{Introduction}

Wilson loop averages are the main observables in Yang-Mills theories,
and it is very important to understand the equations they satisfy.
Best studied are the Ward identities, i.e. direct implications of quantized equations of motion,
which in the case of Wilson averages are known under the  name of  loop equations
\cite{loopeqs,virco,UFN3}  (in the case of the eigenvalue matrix models, they are nothing but the
celebrated Virasoro and W-constraints).
However, there is also a complementary set of relations reflecting the
representation theory background behind the Yang-Mills theory,
and these are studied much worse
because they are associated with Wilson loops in non-fundamental representations,
which are only starting to attract attention within generic Yang-Mills context
\cite{repsinYM}.
The biggest progress on this way is achieved in the $3d$ Chern-Simons theory \cite{CS},
which is topological, and hence does not possess any dynamical loop equations
so that the pure representation theory equations are the only ones which remain.
Wilson loop averages in this case are known as {\it knot polynomials} \cite{knotpols},
and equations in question are called {\it non-commutative (quantum) ${\cal A}$-polynomials} \cite{Gar}.

The ${\cal A}$-polynomials have a number of applications, basically to and around Kashaev large-$r$ limit
\cite{Kash,Kashmore}.
However, this is only a beginning of the story.
Being {\it exactly} calculable non-perturbative quantities,
both the HOMFLY-PT polynomials and ${\cal A}$-polynomials
provide a nice possibility of studying the basic questions in quantum field theory,
from integrability \cite{MMM1} to topological expansions \cite{LMOV}
or the resurgent structure \cite{GaMa}.

In fact, the ${\cal A}$-polynomials appear to be very complicated and difficult to study in any systematic way.
The way out was proposed almost from the very beginning in \cite{Gar05}:
it suggests to look not at the knot polynomials themselves, but at what is now known
as the coefficients of their differential expansion (DE) \cite{DGR,evo,GGS,arthdiff,BiM}.
The quantum {\it ${\cal C}$-polynomials} made out of these coefficients in a more
direct way than the usual knot polynomials (Wilson averages) were first studied
in \cite{Gar05} but then attracted surprisingly low attention
(most probably because of a limited knowledge abut the DE coefficients and the lacking relation to other popular subjects though their classical limit is related by a rational map with the classical ${\cal A}$-polynomials). {\bf In this paper, we return to the abandoned story of ${\cal C}$-polynomials and consider them in the general case of HOMFLY-PT polynomials.}
The original considerations \cite{Gar,Guk,Gar05} were mostly restricted to the Jones polynomials, which are functions only of the parameter $q=e^{2\pi i/(\kappa+2)}$, $\kappa$ being the Chern-Simons coupling constant, and of the representation, and describe Wilson averages in the Chern-Simons theory with the gauge group $SU(2)$. On the contrary, the HOMFLY-PT polynomials are functions of parameters $q=e^{2\pi i/(\kappa+N)}$, $A=q^N$, and of the representation, and describe Wilson averages in the Chern-Simons theory with the gauge group $SU(N)$. Because of an additional parameter $A$, there emerge additional difference equations with respect to shifts of $A$ (additional ${\cal C}$-polynomials), and we discuss the full ring of these ${\cal C}$-polynomials,
which opens new approaches and perspectives.
Being polynomials of $A$ and $q$, HOMFLY-PT are easily analytically continued to arbitrary complex
values of these parameters, and we do not address here the interesting subject of
specific representations arising when they are roots of unity,
in application to $C$-polynomials this will be considered elsewhere.

The original quantum ${\cal A}$-polynomial \cite{Gar} was introduced to describe a difference equation w.r.t. representation label
for the unreduced Jones polynomials of knot ${\cal K}$
\be\label{cycl}
J_{[r]}^{\cal K} = \sum_{n=0}^r \frac{[r+n+1]!}{[r-n]!}\cdot \{q\}^{2n} \cdot
\left.F_{[n]}^{\cal K}(A,q)\right|_{A=q^2}
\ee
where $F_{[n]}$ are the coefficients of differential expansion of the unreduced HOMFLY-PT polynomials colored with symmetric representation $[r]$,
\be
H^{\cal K}_{[r]}(A,q) = \sum_{n=0}^r Z_{[r]}^{[n]}(A,q)\cdot F^{\cal K}_{[n]}(A,q)\nn\\
Z_{[r]}^{[n]}(A,q):=Bin_q(r,n)\cdot\prod_{i=1}^n\{Aq^{r+i-1}\}\cdot\{Aq^{i-2}\}
\label{DEhomfly}
\ee
at the specialization to the Jones polynomials $A=q^2$. The notation used in these formulas can be found at the end of the Introduction.

This linear transformation of the Jones polynomial, (\ref{cycl}) was first considered in \cite{Hab} at $q$ being roots of unity, hence the name for the DE coefficients cyclotomic function.
The knot-independent $Z$-factors for symmetric representations
were found in \cite{IMMMfe} and \cite{evo}
(they are actually known for all rectangular representations for defect-zero knots
\cite{M16,KNTZ},
for the situation with other defects and non-rectangular case, see \cite{BiM} and references therein).

Less known than the ${\cal A}$-polynomials, the quantum ${\cal C}$-polynomials \cite{Gar05}\footnote{S. Garoufalidis et al. introduced ${\cal A}$-polynomials \cite{Gar} and ${\cal C}$-polynomials \cite{Gar05}. We do not know what the name ``${\cal B}$-polynomials" was reserved for, but clearly the properties of knot polynomials is rich enough to find an appropriate candidate.}
describe equations for the DE coefficients $F_{[n]}$ themselves
and are not restricted to the Jones polynomials.
Instead they should be formulated differently for knots with different defects \cite{Konodef}
(which are defined by the powers of the corresponding Alexander polynomials),
because the factorization properties of the differential expansion depends on it.
With the increase of defect, the $Z$-factors get smaller
(and are halved in the limit of large defects),
while the DE coefficients become more involved and acquire additional sub-structures \cite{arthdiff},
which will probably show up at the level of ${\cal C}$-polynomials as well.
In this paper we concentrate on the examples of defect-zero knots only and,
more important, on symmetric representations, i.e. on representations associated with the single-line Young diagrams $[r]$.

In \cite{arthdiff}, we suggested that the DE coefficients can be a better characteristic of knot
than the original knot polynomials.
They have no direct interpretation in Chern-Simons theory and can instead be a base for alternative
effective theory of knots, of which the examples were attempted in \cite{arbor,families} and \cite{tangles}.
This direction seems also related to the Khovanov calculus \cite{Kho} and its far-going generalizations
\cite{Khogen}.
Unfortunately, study of the DE coefficients remains very difficult,
it is still based more on particular examples and computer experiments.
This is true for quantum ${\cal C}$-polynomials as well:
the search for them remains more a piece of art than a regular method.

The symmetric DE coefficients $F_{[n]}^{\cal K}(A,q)$ for a given knot ${\cal K}$
are functions of three variables $A$, $q$, and $n$, and therefore would satisfy three different equations.
However, we restrict ourselves to two equations and treat $q$ as {\it a parameter} rather than {\it a variable}.
As usual, each equation can be considered as a {\it quantum polynomial},
and, for every knot, the two non-commuting quantum polynomials generate
a non-trivial algebra with operator-valued
structure ``constants",
which is the main object of our study in this paper.
It is quite intriguing that such simple quantities as the DE coefficients
for the twist knots are associated with highly non-trivial quantum polynomials and
algebras.

The paper is organized as follows. In section 2, we discuss general existing equations for knot invariants. In section 3, we consider first examples of the equations for a few first terms of the family of twist knots. In section 4, we continue considering examples with a discussion of rings that are formed by ${\cal C}$-polynomials. In section 5, we describe the ${\cal C}$-polynomials for the entire twist knot family, while, in section 6, obtain the ${\cal C}$-polynomials in the simplest example of non-twist (and non-double-braid) knot of still zero defect, knot $9_{46}$. Section 7 contains some concluding remarks. The most tedious formulas are collected in two Appendices.

\paragraph{Notation.} First of all, all knot polynomials are unreduced and given in the topological framing.

As soon as throughout the paper we deal with the symmetric representations only, we denote the DE coefficients instead of $F_{[n]}$ just as $F_n$.

The quantum numbers are defined as $[m]:=\frac{\{q^m\}}{\{q\}}=q^{1-m}[[m]]$ with $\{z\}:=z-z^{-1}$\\
and $[[m]]=1+q^2+\ldots + q^{2(m-1)}$.

The $q$-binomial coefficients are denoted as $Bin_q(n,k)={[n]!\over [k]![n-k]!}$.

With each representation $\mu$, we associate the Young diagram $\mu:\ \mu_1\ge\mu_2\ge\ldots\ge\mu_{l}> 0$, and $|\mu|:=\sum_{i=1}^l\mu_i$, where $l$ is the number of lines in the Young diagram $\mu$, which we often denote $l_{_\mu}$.

\section{Equations on knot polynomials: generalities}

\subsection{Motivation: from equations on characters to ${\cal C}$-polynomials}

We now present one more a kind of motivation for studying $A$ and ${\cal C}$-polynomials,
which is often ignored in the literature.

The irreducible representation of $SL_N$ group associated with the Young diagram $R$ is described by the character $\chi_{R}\{p\}$, which depends on traces of powers of the $SL_N$ group element in the fundamental representation, $p_n=\Tr g^n$ and is the Schur polynomial.  These characters form a closed algebra
\be\label{chex}
\chi_{R_1}\{p\}\cdot \chi_{R_2}\{p\} = \sum_{R\in R_1\otimes R_2} N_{R_1R_2}^R\cdot \chi_R\{p\}
\ee
where $N_{R_1R_2}^R$ are known as Littlewood-Richardson coefficients or structure constants.
Alternatively, one can say that each $\chi_R $ induces a linear operator
acting on characters:
\be
\hat {\cal O}_R : \ \ \ \chi_{Q}  \longrightarrow \sum_{Q'\in R\otimes Q} N_{RQ}^{Q'}\cdot \chi_{Q'}
\ee
If the character variables are restricted to the topological locus
\be
p_k^*:=\frac{\{A^k\}}{\{q^k\}}=\frac{[Nk]}{[k]}
\ee
with $A:=q^N$, one gets an action on the space of quantum dimensions
$D_R(N) := \chi_R\left\{p_k^* \right\}$.
If we further restrict ourselves to $N=2$ (i.e. $SL_2$), we get an action on the space of symmetric representations only,
\be
\hat {\cal O}_{[r]}: \ \ \ D_{[n]}(2) \longrightarrow \sum_{m=0}^{r} D_{[n-r+2m]}(2)
\ee
Moving all the terms in (\ref{chex}) to the l.h.s., one obtains a system of linear equations for characters:
an equation per each $R$.
The simplest of them, associated with $R=[1]$ is just
\be
(q+q^{-1})D_{[n]}(2) - D_{[n+1]}(2)-D_{[n-1]}(2) = 0,
 \ \ \ \ \  {\rm i.e.}\ \ \ \ \   (q+q^{-1})[n+1] - [n+2]-[n]=0
\ee

Knot polynomials are non-trivial deformations of characters (or, at least, of quantum dimensions)
\cite{MMM1}, moreover,
they provide an entire {\it family} of deformations labeled by knots (or, perhaps, better to say, by braids).
As time goes we find more and more demonstration of this analogy,
which forces us to take it seriously.
In particular, quite some time ago, the colored Jones polynomials (the colored HOMFLY-PT polynomials at $N=2$) were demonstrated
to satisfy some linear equations \cite{Guk} nicknamed non-commutative (quantum) ${\cal A}$-polynomials \cite{Gar},
which can be thought of some analogy of the above linear equations for characters.
It was also realized that the equations actually simplify if written not on the knot polynomials
themselves, but on the coefficients of their differential (also called cyclotomic) expansion,
these equations were nicknamed ${\cal C}$-polynomials \cite{Gar05}.

An extension to the generic HOMFLY-PT case is pretty difficult, because, as in the case of characters,
the equations involve not only symmetric representations but also other Young diagrams
with no more than $N-1$ lines. However, it still can be studied in the case of HOMFLY-PT polynomials colored with symmetric representations for particular knots \cite{IMMMfe,Guk2}. Similarly to the $N=2$ case, the ${\cal A}$-polynomials turns out to be much more involved as compared to ${\cal C}$-polynomials. In particular, while the ${\cal A}$-polynomials for the whole series of twist knots are unknown, it is possible to construct ${\cal C}$-polynomials for them, as we demonstrate it in the present paper.

Since the DE coefficients depend on {\it two} variables: $[n]$ (representation) and $A=q^N$, one can actually expect {\it pairs} of equations emerging, and, since they do not necessarily commute,
actually a whole algebra of equations: an algebra of ${\cal C}$-polynomials. The simplest examples of the ${\cal C}$-polynomials and of their algebras is the subject of the present paper.

\subsection{Non-local difference equations}

\paragraph{Knot-independent sum rules.}
We begin with a kind of trivial or, if one prefers, of universal ${\cal C}$-polynomial,
which is almost independent of the knot, and therefore deserve the name of {\it sum rules}.
The point is that the DE coefficients $F_{n}^{\cal K}(A,q)$ are Laurent polynomials in $A^2$ and $q^2$,
and their degree in $A^2$ is often determined by the {\it special} polynomial\footnote{For instance, this is the case for the fundamental HOMFLY-PT polynomial for all 250 knots with up to 10 crossings from the Rolfsen table \cite{katlas} for exception of 19 knots with non-trivial DE coefficient, 6 knots with $F_{1}^{\cal K}(A,q=1) =0$ (knots $8_{14},\ 10_{82},\ 10_{108},\ 10_{116},\ 10_{118},\ 10_{146}$) and 38 knots with $F_{1}^{\cal K}(A,q=1) \sim A^k$ (where is an integer).}, which is the specialization of the HOMFLY-PT polynomial at $q=1$
trivially depending on the representation \cite{DMMSS}:
\be\label{spp}
H^{\cal K}_{R}(A,q=1) = \Big(H^{\cal K}_{[1]}(A,q=1)\Big)^{|R|}
= \Big(1+\{A\}^2F_{1}^{\cal K}(A,q=1)\Big)^{|R|}
\ \ \stackrel{(\ref{DEhomfly})}{\Longrightarrow} \ \
F_{n}^{\cal K}(A,q=1) = \left(F_{1}^{\cal K}(A,q=1)\right)^n
\ee
If $F_{1}^{\cal K}(A,q) = \sum_{m=m_1}^{m_2} c_m(q)A^{2m}$, then
$F_{n}^{\cal K}(A,q=1)$ has powers of $A^2$ in between $nm_1$ and $nm_2$.

Now, since
\be
\sum_{j=0}^{M+1}  {(-x)^j} \cdot\frac{[M+1]!}{[j]![M+1-j]!} =
\prod_{i=0}^M (q^{M-2i}-x)
\label{qubinom}
\ee
the sum
\be
\sum_{j=0}^{M+1}  \frac{{(-q^{2m})^j}}{q^{(2p+1)Mj}} \cdot\frac{[M+1]!}{[j]![M+1-j]!} =
\prod_{i=0}^M \Big(q^{M-2i}-q^{2m-(2p+1)M}\Big)
\ee
vanishes for all $m$ in between $nm_1=pM$ and $nm_2=(p+1)M$.
These boundaries coincide with $nm_1$ and $nm_2$ when
$M=n(m_2-m_1)$ and $2p+1 = \frac{m_2+m_1}{m_2-m_1}:=\nu$.
This means that whenever $\nu$ is odd integer\footnote{This property is also often fulfilled for knots. For instance, this is the case for all knots with 5 and 6 crossings from the Rolfsen table \cite{katlas}, for 2 knots with 7 crossings, etc. In particular, this is the case for
twist knots with negative twist, when $\nu=-1$, while for the positive twist $k$ $\nu={k+1\over k-1}$. For a generic knot, $\nu$ is not odd, and is not even integer.},
\be
\frac{m_2+m_1}{m_2-m_1} \in \mathbb{Z}_{\rm odd} \ \ \Longrightarrow \ \ \
\boxed{
\sum_{j=0}^{M+1}  \frac{{(-)^j}}{q^{(m_2+m_1) nj}}
\cdot\frac{[M+1]!}{[j]![M+1-j]!}\cdot
F_{n}(q^j\!A,q) = 0
}
\ \ \ \ {\rm with} \ \ M=n(m_1-m_2)
\label{sumrule1}
\ee
The number of items in these sum rules depends on representation size $n$,
we call such equations {\it non-local}. An important feature of this kind of equation is that
it does not depend on the knot, but only on the minimal and maximal degrees of its special polynomial.

\paragraph{Knot-dependent semi-local equations.}
We also distinguish {\it semi-local} equations, which are certainly non-local since the numbers of items in them
grow with $n$, but slower than $n$, say, like $n/2$:
these are already restrictive but not too much as compared to explicit expressions for $F_n$. However, the semi-local equations essentially depends on knot details. We will consider examples of the both types of non-local equations in the next sections.

\subsection{Local difference equations: ${\cal C}$-polynomials}

More interesting are {\it local} relations, where the number of terms does not depend on $n$. There are basically two types of local equations: those involving items at different $n$ at the same $A$ (we call them $n$-evolution equations), and those involving shifted $A$ at the same $n$ (we call them $A$-evolution equations). Having these two types of equations, one can certainly generate many other local equations of a mixed type with both $n$- and $A$-evolution presented. However, all these new equations are corollaries of the basic two.

\paragraph{$n$-evolution equations.} The $n$-evolution equations were first considered in \cite{Guk,Gar} in the case of Jones polynomials. In \cite{Gar}, it was proposed that a proper classical limit of these equations (for the Jones polynomials) coincide with the classical ${\cal A}$-polynomial (the hypothesis known as AJ-conjecture). This is why these equations have been called non-commutative ${\cal A}$-polynomials. Their generalization to equations for the HOMFLY-PT polynomials colored with symmetric representations (and further to the superpolynomials) were considered in \cite{Guk2,Gar3}.

\paragraph{$A$-evolution equations.} In contrast with the $n$-evolution, the $A$-evolution equations were first considered later \cite{IMMMfe}, since they require changing $A$, and, hence, cannot be formulated for the Jones polynomials, only for HOMFLY-PT ones, while the first example of the generic symmetric HOMFLY-PT polynomial was obtained in \cite{IMMMfe}. In \cite{IMMMfe}, there is only mixed evolution presented, and, in this paper, we present the pure $A$-evolution.

In the next section, we demonstrate all these types of equations in the simplest case of first twist knots, and, in section 6, in example of a simple non-twist knot $9_{46}$.

\subsection{Genus expansion and the special point}

Expansions of the HOMFLY polynomials can be performed in a few different ways: one can parameterize $q:=e^\hbar$, $A:=e^{N\hbar}$ and consider the limit of $\hbar\to 0$ (such an expansion gives rise to the Vassiliev invariants \cite{Vas}), one can alternatively keep $A$ fixed in the same limit. This expansion is called genus expansion \cite{MMS}, we keep the representation in this limit fixed. When the size of the representation infinitely grows consistently with $\hbar^{-1}$, one obtains the volume conjecture limit \cite{Kash} of the Jones polynomial, and, for the equations, this limit gives rise to either the classical ${\cal A}$-polynomial \cite{Gar}, or to the augmentation variety \cite{AENV}, depending on details.

The genus expansion can be performed both in equations and in the polynomials.
In particular, the leading term of expansion of the polynomial is, as follows from (\ref{spp}),
\be
\Big(1+F_1^{\cal K}(q=1)\{A\}^2\Big)^r = \sum_{n} F_n^{\cal K}(q=1)\cdot \frac{r!}{n!(r-n)!}\{A\}^{2n}
\ee
which explains a natural appearance of binomial coefficients in their explicit expressions.
However, the quantization, i.e. switching on of $q\neq 1$, or taking into account higher genera can be quite nontrivial.

The genus expansion of equations will be discussed later.

\section{${\cal C}$-polynomials for twist knots: examples
\label{Exa}}

In this section, we provide some details about the DE coefficients for first few knots of the twisted family.
This is the case when the DE coefficients are best studied, even far beyond the symmetric representations
but still not in full generality \cite{evo,Mo,BiM}.
It is especially important because it provides exclusive Racah matrices $\bar S$ and $S$,
needed for arborescent calculus and its further generalizations \cite{arbor,families}.

\subsection{DE coefficients for twist and double-braid knots}

For the twist knots, there is a general formula for symmetric DE coefficients \cite{evo}:
\be
F^{{\rm twist}_k}_{n}=q^{n(n-1)/2}A^n \sum_{j=0}^n
(-)^j\frac{[n]!}{[j]![n-j]!}\frac{ \{Aq^{2j-1}\}\cdot(Aq^{j-1})^{2jk}}{\prod_{i=j-1}^{n+j-1}\{Aq^i\}}
\ee
It follows from evolution in $k$ and can be rewritten as
\be
F^{{\rm twist}_k}_{n} = \sum_{m} ({\cal B}^{k+1})_{[n],[m]}
\ee
i.e. made from elements of the KNTZ matrix \cite{KNTZ}
\be
{\cal B}^{-1}_{\lambda,\mu} =
\frac{\chi_{\lambda/\mu}^{(0)}\chi_\mu^{(0)}}{\chi_\lambda^{(0)}}
\cdot \Lambda_\mu^{-1}
\ee
evaluated at the locus $p^{(0)}_k = \frac{1}{\{q^k\}}$.
$A$ enters entirely through $\Lambda_\mu = \Big(A^{|\mu|}q^{\varkappa(\mu)}\Big)^2$, which is proportional to the square of eigenvalue of the corresponding $R$-matrix. Here $\varkappa(\mu):=\sum_{i=1}^{l_{_\mu}}\mu_i(\mu_i-2i+1)$.

One of the properties of the twist knot family is that their DE coefficients for positive and negative twists are related by a simple formula:
\be\label{kd1}
{F^{{\rm twist}_{k}}_{n}(A,q)\over F^{{\rm twist}_{1}}_{n}(A,q)} = {F^{{\rm twist}_{-k}}_{n}(A^{-1},q^{-1})
\over F^{{\rm twist}_{-1}}_{n}(A^{-1},q^{-1})}
\ee
Since
\be
F^{4_1}_{n} = F^{{\rm twist}_{-1}}_{n} = 1, &\ \ \ \ &
F^{3_1}_{n} = F^{{\rm trefoil}}_{n} = F^{{\rm twist}_{1}}_{n} = (-)^nA^{2n}q^{n(n-1)}
\ee
formula (\ref{kd1}) becomes
\be\label{kd}
\boxed{
F^{{\rm twist}_{k}}_{n}(A,q)= (-)^nA^{2n}q^{n(n-1)}\cdot F^{{\rm twist}_{-k}}_{n}(A^{-1},q^{-1})}
\ee

From any of these expressions, the DE coefficients are easily calculated
and described in the third way as multiple $q$-binomial sums. Indeed, introduce a function
\be
{\cal P}_1(n)  = \sum_{j=0}^n   A^{-2j}q^{-j(j+n-2)} \frac{[n]!}{[n-j]![j]!}
\ee
and its iterations
\be
{\cal P}_m(n)   = \sum_{j=0}^n {\cal P}_{m-1}(j)\cdot A^{-2j}q^{-j(j+n-2)}\frac{[n]!}{[n-j]![j]!}
\ee
Then
\be
\boxed{
F^{{\rm twist}_{-k}}_{n} = {\cal P}_{k-1}(n)
}
\ee

%\noindent

\bigskip

For the double braids, there is a mysterious product formula \cite{M16}
\be\label{dd}
F^{{\rm db}_{k,l}}_{n} = \frac{F^{{\rm twist}_k}_{n}F^{{\rm twist}_l}_{n}}
{F^{{\rm trefoil}}_{n}}
\ee
It would be interesting to understand what this means for the ${\cal C}$-polynomials.

\subsection{${\rm twist}_{-1}=4_1$}

For the figure-eight knot $4_1$, all the DE coefficients
$F_{n}^{{\rm twist}_{-1}}(A,q) = 1$, and all the equations are trivial.

\subsection{${\rm twist}_{1}=3_1$}

For the trefoil $3_1$, the DE coefficients are not quite unities, still they are simple
monomials:
\be
F_{n}^{{\rm twist}_{1}}(A,q) = (-)^nA^{2n}q^{n(n-1)}
\ee
hence, the $n$- and $A$-evolution equations are rather trivial:
\be
F_{n+1}^{{\rm twist}_{1}}(A,q) = -A^2q^{2n}\cdot F_{n}^{{\rm twist}_{1}}(A,q)
\nn \\
F_{n}^{{\rm twist}_{1}}(Aq,q) = q^{2n}F_{n}^{{\rm twist}_{1}}(A,q)
\ee

\subsection{${\rm twist}_{-2}=6_1$
\label{61base}}

Truly non-trivial equations emerge for the first time in the case of $6_1$ knot,
where
\be
F_{n}^{{\rm twist}_{-2}}(A,q) = \sum_{j=0}^n\frac{[n]!}{[j]![n-j]!}\cdot\frac{1}{A^{2j}q^{j(n+j-2)}}
=\sum_{j=0}^n\frac{
[[n]]! }{[[j]]![[n-j]]!}\cdot\frac{1}{(Aq^{n-1})^{2j} }
\label{Ftw-2}
\ee
\noindent
To avoid possible confusions from now on, we {\bf omit the argument} $q$ of $F_n(A,q)$,
when it does not matter
(is the same in all the terms of equation).

As we discussed in section 2, the DE coefficients ${F}_{n}^{{\rm twist}_{-2}}(A,q)$
satisfy a set of local and non-local difference equations:

\begin{itemize}

\item[\underline{$n$-evolution:}]
\be
\!\!\!\!\!\!\!\!\!\!\!
\boxed{
 {F}_{n+1}^{{\rm twist}_{-2}}(A) - \Big(1+\frac{1}{A^2q^{2n}}\Big)F_{n}^{{\rm twist}_{-2}}(A)
=q^{-2}(1-q^{-2n})
\left(\Big( 1- \frac{1}{A^2q^{2n-2}}\Big) \cdot F_{n}^{{\rm twist}_{-2}}(A)
 - {F}_{n-1}^{{\rm twist}_{-2}}(A)\right)
 }
\label{61n}
\ee

\item[\underline{$A$-evolution:}]

\be
\boxed{
F_{n}^{{\rm twist}_{-2}}(A) = \left(1-\frac{1}{A^2q^{2n}}\right)F_{n}^{{\rm twist}_{-2}}(Aq)
+\frac{1}{A^2}F_{n}^{{\rm twist}_{-2}}(Aq^2)
}
\label{61A}
\ee
\end{itemize}
In the genus expansion $q=e^h$ with $h\longrightarrow 0$, the first equation is reduced to $F_{n+1}^{{\rm twist}_{-2}}=
\left(1+\frac{1}{A^2}\right)\cdot F_{n}^{{\rm twist}_{-2}}$, while the second one implies
$A\frac{\p F_{n}^{{\rm twist}_{-2}}}{\p A} = -2n\frac{F_{n}^{{\rm twist}_{-2}}}{A^2+1}$.
This  is obviously true for arbitrary $n$ for the special polynomial
$\left.F_{n}^{{\rm twist}_{-2}}\right|_{q=1}
=\Big(\left.F_{1}^{{\rm twist}_{-2}}\right|_{q=1}\Big)^n
= \left(1+\frac{1}{A^2}\right)^n$, which is, indeed, the case for $6_1$ knot.

\begin{itemize}
\item[\underline{Non-local eq.}]
As follows from (\ref{Ftw-2}), one has to choose $m_1=-1$ and $m_2=0$ in (\ref{sumrule1}) so that the non-local sum rule is
\be
\sum_{j=0}^{n+1} (-)^jq^{nj}\cdot \frac{[n+1]!}{[j]![n+1-j]!}\cdot
F_{n}^{{\rm twist}_{-2}}(Aq^j) = 0
\label{sumruletwist-2}
\ee

\item[\underline{Semi-local eq.}] In order to obtain a semi-local equation, one has to combine the $A$-shifts with $n$-shifts (otherwise, the equations are the above local equations). The leading dependence of $F_n$ on $n$ is typically
$\sim A^{\alpha n}q^{\beta n^2}$, i.e. the power of $A$ is linear in $n$,
while that of $q$ is quadratic in $n$.
This means that an infinitesimal/continuous equation would be of the diffusion type
$\p/\p\log A \sim \p^2/\p(\log q)^2$.
A finite-difference lift of the diffusion equation is more involved.
Still, such equations exist.
Looking at the last representation in (\ref{Ftw-2}), it is clear that one could make
a $q$-shift $q\longrightarrow q^2$ and compensate it by an $A$-shift $A\longrightarrow q^{1-n}A$ modulo
the change of binomial coefficient $\frac{[[n]]!}{[[j]]![[n-j]]!}$.
This change, however, is absent for the ``boundary" terms with $j=0$ and $j=n$,
which leads to
\be
F_n^{{\rm twist}_{-2}}(Aq^{1-n},q^2) - F_n^{{\rm twist}_{-2}}(A,q) = \sum_{j=1}^{n-1}
\left(\left.\frac{[[n]]!}{[[j]]![[n-j]]!}\right|_{q=q^2}
- \left.\frac{[[n]]!}{[[j]]![[n-j]]!}\right|_{q=q}\right)\cdot
\frac{1}{(Aq^{n-1})^{2j}}
\ee
with the sum at the r.h.s. can be expressed through $F_{n'}^{{\rm twist}_{-2}} $ with $n'\leq n-2$.
After some algebra, one can finally obtain a semi-local equation
\be
\boxed{
F_n^{{\rm twist}_{-2}}(A,q) =
\sum_{j=0}^{n/2} \frac{(-)^j(q^2-1)^j}{A^{2j}q^{3j(j-1)}} \frac{[n]!}{[n-2j]![2j]!!}\cdot
F_{n-2j}^{{\rm twist}_{-2}}(Aq^{4j+1-n},q^2)
}
\label{61q}
\ee
It contains an $n$-dependent sum, but it is two times shorter than $n$, and hence is
a restrictive relation for the DE coefficients $F_n$.
Note that the double factorial appears in denominator so that the weight in the sum
deviates from the ordinary $q$-binomial coefficient.
\end{itemize}

\bigskip

\noindent
The local pair (\ref{61n}), (\ref{61A})
provides a set of independent equations for the particular case of the knot $6_1$.
In addition, there are plenty of other {\it local} equations where both $n$ and $A$ are shifted.
It is natural to assume they are all corollaries of the first two,
which is confirmed in sec.\ref{algebra}.
Note that we deal only with two equations (\ref{61n}) and (\ref{61A}), though the DE coefficients depends on the three parameters $n$, $A$ and $q$: it seems more natural to treat $q$, which specifies the $A$-shift
(or, perhaps better, the $N$-shift)
as a parameter rather than an independent third variable.
This is also justified by the fact that the $q$-shifts satisfy only {\it semi-local} equations.

\subsection{${\rm twist}_{2}=5_1$ and $k\longleftrightarrow -k$ duality}

The DE coefficients themselves are analytical functions of $k$
described by the same formulas at positive and negative $k$.
This is a manifestation of $k\longleftrightarrow -k$ duality for the twist family.
However, the equations are slightly different for positive and negative $k$, (\ref{kd}).
This can be a sign of their relation to superpolynomials
and Khovanov calculus \cite{Kho}
which are also known to slightly break $k\longleftrightarrow -k$
symmetry \cite{evo,sats,AMP}.

In particular, in the case of knot
${\rm twist}_{2}=5_2$:
\be
\!\!\!\!\!\!\!\!\!\!\!
F_{n+1}^{{\rm twist}_{2}}(A,q) + (q^{2n}+A^2q^{6n})\cdot F_{n}^{{\rm twist}_{2}}(A,q)
= (q^{2n}-1)\Big(A^2q^{2n+2}F_{n}^{{\rm twist}_{2}}(A,q)+q^{4n}F_{n-1}^{{\rm twist}_{2}}(A,q)\Big)
\nn \\
F_{n}^{{\rm twist}_{2}}(A,q) = \left(\frac{1}{q^{2n}}-A^2\right)F_{n}^{{\rm twist}_{2}}(Aq,q)
+\frac{A^2}{q^{4n}}F_{n}^{{\rm twist}_{2}}(Aq^2,q)
\ee
Since $F_{n}^{{\rm twist}_{2}}(A,q)=A^2(1+A^2)$,
the sum rule (\ref{sumrule1}) is
\be
\sum_{j=0}^{n+1} \frac{(-)^j}{q^{3nj}}\cdot \frac{[n+1]!}{[j]![n+1-j]!}\cdot
F_{n}^{{\rm twist}_{2}}(Aq^j) = 0
\label{sumruletwist2}
\ee

\subsection{${\rm twist}_{-3}=8_1$
\label{81main}}

As an intermediate step from sec.3 to generic formulas, we
list explicit expressions for one more twist knot, ${\rm twist}_{-3}=8_1$.

\begin{itemize}

\item[\underline{$n$-evolution:}]
\be
\boxed{
\begin{array}{c}
  {F}_{n+1}^{{\rm twist}_{-3}}(A)
  - \Big(1+\frac{1}{A^2q^{2n}}+\frac{1}{A^4q^{4n}}\Big)\cdot F_{n}^{{\rm twist}_{-3}}(A)
  =   \\ \\
=q^{-3}(1-q^{-2n})\left\{
\Big([2]-\frac{q^{2-3n}(q^{n+1}-q^{-n-1})}{A^2} - \frac{q^{3-5n}(q^n+q^{-n})}{A^4}\Big)
\cdot F_{n}^{{\rm twist}_{-3}}(A)
+ \right.  \\ \\ \left.
+ \Big(-q^{-1}[3]+q^{-2n-1}-\frac{[2]}{q^{4n-2}A^2}\Big)\cdot F_{n-1}^{{\rm twist}_{-3}}(A)
+ q^{-3}(1-q^{2-2n})\cdot F_{n-2}^{{\rm twist}_{-3}}(A)
\right\}
\end{array}
}
\ee

\item[\underline{$A$-evolution:}]
\be
\boxed{
\begin{array}{c}
\Big(q^{2n+2}A^2-1\Big)\cdot A^6q^{4n+2}\cdot F_{n}^{{\rm twist}_{-3}}(A)-\\ \\
\Big(q^{2n}A^2-1\Big)\Big(A^2q^{2n+2}+q^{2n+2}-1\Big)
\cdot A^4q^{2n+2}\cdot F_{n}^{{\rm twist}_{-3}}(Aq)-\\ \\
-\Big(q^{2n+2}A^2-1\Big)\Big(1+q^{2n}A^2(q^{2n+2}-1)\Big)\cdot F_{n}^{{\rm twist}_{-3}}(Aq^2)+
\\ \\
+\Big(q^{2n}A^2-1\Big)\cdot q^{4n}\cdot F_{n}^{{\rm twist}_{-3}}(Aq^3) =0
\end{array}
}
\ee

\item[\underline{Non-local eq.:}]
As usual, there is a knot-independent non-local sum rule (\ref{sumrule1}),
which, in this case, reads
\be
\sum_{j=0}^{2n+1} (-)^jq^{2nj}\cdot \frac{[2n+1]!}{[j]![2n+1-j]!}\cdot
F_{n}^{{\rm twist}_{-3}}(Aq^j) = 0
\label{sumruletwist-3}
\ee
\end{itemize}

\section{Algebra of ${\cal C}$-polynomials
\label{algebra}}

\paragraph{Twist -1 knot.} Introduce two operators $\hat\Delta_n$, $\hat D_A$,
acting on any function $f_n(A)$ of two variables by the rules
\be
\hat\Delta_n\cdot f_n(A)=f_{n+1}(A)  \nn \\
\hat D_A\cdot f_n(A)=f_n(Aq) \
\ee
Then the equations for $F_n^{{\rm twist}_{-1}}(A)=1$
are just
\be
\left\{\begin{array}{c}
(\hat\Delta_n-1)\cdot F_{n}^{{\rm twist}_{-1}}(A)=0\\
(\hat D_A-1)\cdot F_{n}^{{\rm twist}_{-1}}(A)=0
\end{array}
\right.
\ee
and the two operators $\hat{\cal O}_1^{{\rm twist}_{-1}} :=(\hat\Delta_n-1)$,
$\ \ \hat{\cal O}_2^{{\rm twist}_{-1}} :=(\hat D_A-1)$ obviously commute,
\be
\left[\hat{\cal O}_1^{{\rm twist}_{-1}} ,\ \hat{\cal O}_2^{{\rm twist}_{-1}}\right]=0
\label{O1O2twist-1}
\ee

\paragraph{Twist -2 knot.} Similarly,
the equations (\ref{61n}) and (\ref{61A}) for twist $-2$ can be rewritten as
\be
\left\{\begin{array}{c}
\widehat{\cal O}_1^{{\rm twist}_{-2}}\cdot F_{n}^{{\rm twist}_{-2}}(A)=0\\
\widehat{\cal O}_2^{{\rm twist}_{-2}}\cdot F_{n}^{{\rm twist}_{-2}}(A)=0
\end{array}
\right.
\ee
\be
\widehat{\cal O}_1^{{\rm twist}_{-2}}&:=&\hat\Delta_n^2-\left[\left(1+{1\over A^2q^{2n+2}}\right) +{1\over q^2}\left(1-{1\over q^{2n+2}}\right)
\left(1-{1\over A^2q^{2n}}\right)\right]\cdot\hat\Delta_n+{1\over q^2}\left(1-{1\over q^{2n+2}}\right)\\
\widehat{\cal O}_2^{{\rm twist}_{-2}}&:=&1-\left(1-{1\over A^2q^{2n}}\right)\cdot \hat D_A-{1\over A^2}\cdot\hat D_A^2
\ee
The commutator of these operators is
\be
\left[\widehat{\cal O}_1^{{\rm twist}_{-2}},\widehat{\cal O}_2^{{\rm twist}_{-2}}\right]
=[2]{\{q\}\over A^2q^{2n+2}}\cdot
\hat\Delta_n\cdot\hat D_A\cdot\left(\hat\Delta_n-1-{1\over A^2q^{2n}}\hat D_A\right)
=-[2]{\{q\}\over A^2q^{2n+2}}\cdot
\hat\Delta_n\cdot\hat D_A\cdot\widehat{\cal O}_3^{{\rm twist}_{-2}}
\ee
Thus, we generated a new operator
\be\label{O3}
\widehat{\cal O}_3^{{\rm twist}_{-2}}
:= -\frac{A^2q^{2n-2}}{\{q^2\}}\cdot\hat\Delta_n^{-1}\cdot\hat D_A^{-1}\cdot
\left[\widehat{\cal O}_1^{{\rm twist}_{-2}},\widehat{\cal O}_2^{{\rm twist}_{-2}}\right]
=\hat\Delta_n-1-{1\over A^2q^{2n}}\hat D_A
\ee
Its commutation relations
\be
\left[\widehat{\cal O}_1^{{\rm twist}_{-2}},\widehat{\cal O}_3^{{\rm twist}_{-2}}\right]&
=&{\{q\}\over A^2q^{2n+6}}\hat\Delta_n\cdot
\left(\left[qA^2+q^4[2]\hat D_A-{[2]\over q^{2n-2}}\right]\widehat{\cal O}_3^{{\rm twist}_{-2}}
-{[2]\over q^{2n-2}}\widehat{\cal O}_2^{{\rm twist}_{-2}}\right)\nn\\
\left[\widehat{\cal O}_2^{{\rm twist}_{-2}},\widehat{\cal O}_3^{{\rm twist}_{-2}}\right]
&=&{\{q\}\over A^2q^{2n+1}}\hat D_A\cdot \left(\widehat{\cal O}_2^{{\rm twist}_{-2}}+
\widehat{\cal O}_3^{{\rm twist}_{-2}}\right)
\ee
do not involve any new operators,
thus the constraint algebra gets closed.
Everything what one can generate is
a corollary of (\ref{61n}) and (\ref{61A}).

\bigskip

For instance, the newly generated operator $\widehat{\cal O}_3^{{\rm twist}_{-2}}$
gives rise to the equation
\be
F_{n+1}^{{\rm twist}_{-2}}(A) =
F_n^{{\rm twist}_{-2}}(A) + \frac{1}{A^2q^{2n}}F_n^{{\rm twist}_{-2}}(Aq)
\label{62immmfe}
\ee
while the combination
\be
\left({1\over q^{2n}}-1-{1\over q^{2n+2}}\right)\widehat{\cal O}_2^{{\rm twist}_{-2}}
+(\hat D_A-1)\cdot \widehat{\cal O}_3^{{\rm twist}_{-2}}
\ee
produces
\be
F_{n}^{{\rm twist}_{-2}}(A) = F_{n}^{{\rm twist}_{-2}}(Aq)
+ \frac{1-q^{-2n}}{A^2}F_{n-1}^{{\rm twist}_{-2}}(Aq^2)
\label{62Ashift}
\ee
These (\ref{62immmfe}) and (\ref{62Ashift})
imply a whole variety of other {\it local} equations,
where both $n$ and $A$ shifted:

\be\label{46}
F_{n+1}^{{\rm twist}_{-2}}(A) = F_{n}^{{\rm twist}_{-2}}(Aq)
+\frac{1}{A^2}F_{n}^{{\rm twist}_{-2}}(Aq^2)
\ee

\be
F_{n}^{{\rm twist}_{-2}}(Aq^2)-F_{n-1}^{{\rm twist}_{-2}}(Aq^2)
= \frac{1}{q^{2n}}\Big(F_{n}^{{\rm twist}_{-2}}(Aq)-F_{n-1}^{{\rm twist}_{-2}}(Aq^2)\Big)
\ee

\be\label{48}
F_{n+1}^{{\rm twist}_{-2}}(A) =
F_n^{{\rm twist}_{-2}}(A) + \frac{1}{A^2q^{2n}}F_n^{{\rm twist}_{-2}}(Aq^2)
  + \frac{1-q^{-2n}}{A^4q^{2n+2}}F_{n-1}^{{\rm twist}_{-2}}(Aq^3)
\ee
and so on.

The twist $-2$ analogue of the commutativity (\ref{O1O2twist-1}) is now a more sophisticated algebra
of $D$-module constraints
\be
\left[\hat{\cal O}_1^{{\rm twist}_{-2}} ,\
\left[\hat{\cal O}_1^{{\rm twist}_{-2}} ,\ \hat{\cal O}_2^{{\rm twist}_{-2}}\right]\right]
= \hat\alpha_1 \hat{\cal O}_1^{{\rm twist}_{-2}} + \hat\beta_1\hat{\cal O}_2^{{\rm twist}_{-2}}
+ \hat\gamma_1 \left[\hat{\cal O}_1^{{\rm twist}_{-2}} ,\ \hat{\cal O}_2^{{\rm twist}_{-2}}\right]
\nn \\
\left[
\left[\hat{\cal O}_1^{{\rm twist}_{-2}} ,\ \hat{\cal O}_2^{{\rm twist}_{-2}}\right],
\ \hat{\cal O}_2^{{\rm twist}_{-2}} \right]
= \hat\alpha_2 \hat{\cal O}_1^{{\rm twist}_{-2}} + \hat\beta_2\hat{\cal O}_2^{{\rm twist}_{-2}}
+ \hat\gamma_2 \left[\hat{\cal O}_1^{{\rm twist}_{-2}} ,\ \hat{\cal O}_2^{{\rm twist}_{-2}}\right]
\label{O1O2twist-2}
\ee
with operator-valued coefficients (standing at the left of $\hat{\cal O}$).

\paragraph{Twist $-k$ knots for higher $k$.} It is natural to conjecture that, for the twist $-k$,
we will get a closure at the $k$-th commutator level,
but we did not get enough evidence to support it,
despite the form of the ``generators" $\hat{\cal O}_1^{{\rm twist}_{-k}}$
and $ \hat{\cal O}_2^{{\rm twist}_{-k}}$ is explicitly known.

\paragraph{The genus zero limit.}
Consider the genus expansion of algebra at the vicinity of the special polynomial point. As we emphasized after Eqs.(\ref{61n})-(\ref{61A}), in this limit,
\be
\widehat{\cal O}_1^{{\rm twist}_{-2}}&:=&\hat\Delta_n-1-{1\over A^2}\nn \\
\widehat{\cal O}_2^{{\rm twist}_{-2}}&:=&\left(1+{1\over A^2}\right)\cdot {\partial\over\partial A}+{2n\over A^3}
\ee
Commutator of these two operators gives
\be
\left[\widehat{\cal O}_1^{{\rm twist}_{-2}},\widehat{\cal O}_2^{{\rm twist}_{-2}}\right]={2\over A^3}\left(\hat\Delta_n-1-
{1\over A^2}\right)={2\over A^3}\widehat{\cal O}_1^{{\rm twist}_{-2}}
\ee
which is, indeed, consistent with the genus zero part of operator (\ref{O3})
\be
\widehat{\cal O}_3^{{\rm twist}_{-2}}=\hat\Delta_n-1-{1\over A^2}=\widehat{\cal O}_1^{{\rm twist}_{-2}}
\ee
Similarly, in the case of $k=-3$, the two operators in the leading order of the genus expansion are
\be
\widehat{\cal O}_1^{{\rm twist}_{-3}}&:=&\hat\Delta_n-\left(1+{1\over A^2}+{1\over A^4}\right)\nn \\
\widehat{\cal O}_2^{{\rm twist}_{-3}}&:=&\left(1+{1\over A^2}+{1\over A^4}\right)\cdot {\partial\over\partial A}
+{2n\over A^3}\left(1+{2\over A^2}\right)
\ee
so that
\be
\left[\widehat{\cal O}_1^{{\rm twist}_{-3}},\widehat{\cal O}_2^{{\rm twist}_{-3}}\right]=
{2\over A^3}\left(1+{2\over A^2}\right)\widehat{\cal O}_1^{{\rm twist}_{-3}}
\ee
At generic $k\in \mathbb{Z}_{>0}$,
\be
\widehat{\cal O}_1^{{\rm twist}_{-k}}&:=&\hat\Delta_n-{1\over A^{2(k-1)}}{A^{2k}-1\over A^2-1}\nn \\
\widehat{\cal O}_2^{{\rm twist}_{-k}}&:=&{1\over A^{2(k-1)}}{A^{2k}-1\over A^2-1}\cdot {\partial\over\partial A}
+{2n\over A^3}{A^{2(k+1)}+k-A^2(1+k)\over A^{2(k-1)}(1-A^2)^2}
\ee
so that
\be
\left[\widehat{\cal O}_1^{{\rm twist}_{-k}},\widehat{\cal O}_2^{{\rm twist}_{-k}}\right]=
{2\over A^3}{A^{2(k+1)}+k-A^2(1+k)\over A^{2(k-1)}(1-A^2)^2}\widehat{\cal O}_1^{{\rm twist}_{-k}}
\ee
Thus, one can see that, in the limit of $q\to 1$, the algebra of ${\cal C}$-polynomials becomes the very simple triangular Lie algebra, however, as we demonstrated above, its $q$-deformation is quite sophisticated.

\section{General formulas for the entire twist family
\label{gentwist}}

From examples of twist knots in section \ref{Exa}, we come here to the general
structure of twisted ${\cal C}$-polynomials for the entire family.

\subsection{Twist knots: $n$-evolution}

Though equations for positive and negative twists are related by the simple formula (\ref{kd}), here we write down the both cases to compare with the $A=q^2$ answers of \cite{Gar05}. Let $k$ always be positive, then
\be\label{n}
\boxed{
\begin{array}{cc}
&{F}_{n+k}^{{\rm twist}_{-k}}(A,q)=\sum_{j=0}^{k-1}  {(-1)^{j+k+1}\over \{q\}^{j-k+1}}\cdot
\left(V^{(-,k)}_{n,j}-q^{-2(k+n)}\cdot V^{(-,k-1)}_{n+1,j-1}\right)\cdot{F}_{n+j}^{{\rm twist}_{-k}}(A,q)
\\ {\rm with} \\
&V^{(-,k)}_{n,j}:=q^{-{1\over 2}(k-1)(3k+2n)-{1\over 2}j(3j+2k+6n-5)}\cdot
{[n+k-1]!\over[n+j]!}\cdot\sum_{i=0}^j{q^{(4n+2k+2j-3)i}\over A^{2(j-i)}}\cdot{[k-i-1]![k-j+i]!\over [k-j-1]![k-j]![j-i]![i]!}
\\ \\
{\rm  and}
\\ \\
&{F}_{n+k}^{{\rm twist}_{k}}(A,q)=\sum_{j=0}^{k-1}  {(-)^{j+k}\over
\{q\}^{j-k+1}}\cdot
\left(V^{(+,k)}_{n,j}-q^{2(k+n)}\, V^{(+,k-1)}_{n+1,j-1}\right)\cdot
{F}_{n+j}^{{\rm twist}_{k}}(A,q)
\\ {\rm with} \\
&V^{(+,k)}_{n,j}:=q^{{1\over 2}(k-1)(5k+2n)+2kn-{1\over 2}j(3j+2k+6n-3)}\cdot
{[n+k-1]!\over[n+j]!}\cdot\sum_{i=0}^j
\frac{q^{(4n+2k+2j-3)i}}{A^{2(j-i-k)}}\cdot
{[k-i]![k-j+i-1]!\over [k-j-1]![k-j]![j-i]![i]!}
\end{array}
}
\ee

\noindent
The number of items in these relations depend on $k$ (i.e. on the knot),
but not on representation size $n$, therefore these are, indeed, {\it local} difference equations.

\subsection{Twist knots: $A$-evolution}

We gave earlier in sec \ref{Exa} examples of $A$-evolution for $k=-2$ and $k=-3$.
The general structure is
\be\label{A}
\boxed{
\begin{array}{cl}
0=&\sum_{j=0}^{k-1} \epsilon_j G_{n,j}^{(k)}
\cdot F_{n}^{{\rm twist}_{-k}}(Aq^j)
+\epsilon_k\cdot q^{2(k-1)n} \cdot \prod_{i=0}^{\frac{k-3}{2}} \Big(A^2q^{2n+2i}-1\Big)
\cdot F_{n}^{{\rm twist}_{-k}}(Aq^k)
\cr\cr
\hbox{where } \epsilon_j =& \sqrt{2}\cdot\cos\left(\frac{(2j+1)\pi}{4}\right) \hbox{ and}\cr\cr
G^{(k)}_{n,j} =& \Big(A^2q^{2n+2j-2}-1\Big)\cdot\prod_{i=0}^{\ent\left(\frac{j-3}{2}\right)}\Big(A^2q^{2n+2i}-1\Big) \cdot \prod_{i=\ent\left(\frac{k+j}{2}\right)}^{k-2}\Big(A^2q^{2n+2i}-1\Big)\times\cr\cr
&\times A^{k^2-(j+1)k+4\fr\left(\frac{k}{2}\right)\fr\left(\frac{j}{2}\right)}q^{2(k-j-1)n+\sigma_{k,j}}\cdot g^{(k)}_{n,j}
\end{array}}
\ee
Here $g^{(k)}_{n,j}$ is a polynomial\footnote{Its degree in $A^2$ is $\ent\left(\frac{k-1}{2}\right)+
2\fr\left(\frac{j+1}{2}\right)\fr\left(\frac{k+1}{2}\right)$ and in $q^{2n}$ is $j$.} in $A^2$, $q^{2n}$ and $q^2$. Note that, in variance with the $n$-evolution equations (\ref{n}), equations (\ref{A})  form two different series for odd and even $k$, since the polynomials $g^{(k)}_{n,j}$ differ in these series, see examples of these polynomials and of the numerical coefficients $\sigma_k$  in Appendix A.

Equation (\ref{A}) (or (\ref{Ashiftsfirst})-(\ref{Ashiftslast}) in Appendix A)
can be considered as a recursion in $m$ for the coefficients
$F_{n|m}^{{\rm twist}_{-k}}$ in
$F_{n}^{{\rm twist}_{-k}}=\sum_{m\geq 0} F_{n|m}^{{\rm twist}_{-k}}A^{-2m}$.
For instance, (\ref{Ashiftsfirst}) implies that
\be
F_{n|m+1}^{{\rm twist}_{-2}} = \frac{q^{2n-2m}-1}{ q^{2n-2}(q^{2m+2}-1)} \cdot F_{n|m}^{{\rm twist}_{-2}}
\ee

\subsection{Twist knots: mixed $n$-$A$-evolutions}

In fact, equations with mixed $n$-$A$-evolutions like that first considered in \cite{IMMMfe}, for instance, Eqs.(\ref{62Ashift}), (\ref{62immmfe}), (\ref{46})-(\ref{48}), etc. can be also generalized to the whole twist series. For instance, Eq.(\ref{48}) would look like
\be
\boxed{
\begin{array}{c}
F_{n+1}^{{\rm twist}_{-k}}(A) = F_{n}^{{\rm twist}_{-k}}(A)
+ \sum_{j=1}^{k-2} \frac{1}{A^{2j}q^{2n}}F_{n}^{{\rm twist}_{-k}}(Aq)
+ \frac{1-\delta_{k,1}}{A^{2(k-1)}q^{2n}}F_{n}^{{\rm twist}_{-k}}(Aq^2) + \\ \\
+ \sum_{j=1}^{{\rm min}(k-1,n)}
\frac{(-)^{j-1}c^{(k)}_j \{q\}^j}{A^{2kj}q^{(2j+1)(j-1)k-\frac{(j-1)(j-2)}{2}+(2+j)n+2}}
\frac{[n]!}{[n-j]!}
F_{n-j}^{{\rm twist}_{-k}}(Aq^{2j+1})
\end{array}
}
\ee

\noindent
with

\be
\sum_{k=2}^\infty c^{(k)}_j z^k = -\sum_{i=0}^\infty \sum_{k=2i+2}^\infty
\left(\frac{C^{(k)}_{j,i}(q)}{A^{2i}q^{k+2ji-1}}
+ (1-\delta_{k,2i+j+1})\frac{C^{(k)}_{j,i}(q^{-1})}{(Aq^{j+1})^{2(k-j-1)}A^{-2i}q^{-k-2ji+1}}
\right)\cdot z^k
\label{ckjgen}
\ee
\be
\boxed{
C^{(k)}_{j,i}(q) = \frac{[i+j-1]!}{[i]!([j]!)^2}\cdot\frac{[k-i-2]!}{[k-i-j-1]!}
\Big([i][k-2j-i-1]+q^{k-i-1}[j][i-j]\Big)
}
\ee

\section{Another defect zero knot}

Of course, quantum ${\cal C}$-polynomials exist not only for the twist knots.
For illustration, we provide an example of defect-zero \cite{Konodef} knot
$9_{46}$, the smallest one beyond the double-braid family.
For the double braid, the coefficients
of differential expansion are given by formula (\ref{dd}).
For higher defects, the coefficients of differential expansion are somewhat reduced,
and it is interesting to see what would this mean for the ${\cal C}$-polynomials:
in variance with ${\cal A}$-polynomials, their {\it definition} begins to depend on the defect.
We leave these questions for the future, and here
we just demonstrate that our discussion above is in no way restricted to the twist family,
in fact, one can lift it to other families of knots like those introduced in \cite{families}
or \cite{tangles}.

Knot $9_{46}$ is a very simple knot, with $F_1^{9_{46}} = A^2(A^2+1)$.
This first DE coefficient is independent of $q$, which
is the sign of the vanishing defect \cite{Konodef}.
Note it is almost the same as $F^{5_{1}}_1 = F^{{\rm twist}_2}_1=  -A^2(A^2+1)$
or $F^{6_{2}}_1 = F^{{\rm twist}_{-2}}_1=   \frac{A^{2}+1}{A^2}$.
However, in general
$F_n^{9_{46}}(A) = F^{{\rm twist}_{-2}}_n\left(\frac{1}{Aq^{3(n-1)}}\right) +
O\Big([n][n-1][n-2]\{q\}^2\Big)$,
and the actual DE coefficients in this case are rather involved
as compared to (\ref{Ftw-2}):
\be
F^{9_{46}}_n(A)
= A^{2n}\sum_{j=0}^{n/2} (-)^j(q^2-1)^j q^{n^2+(2j-1)n-3j(j+1)}
\frac{[n]!}{[n-2j]![2j]!!} \prod_{i=n-1}^{2n-2j-2}(A^2q^{2i}+1)
\label{F946viaprod}
\ee
Quantum binomial formula (\ref{qubinom}) can be applied to convert the product at the r.h.s.
into another sum, but (\ref{F946viaprod}) itself
contains double factorial and is similar in this respect to (\ref{61q}).

\bigskip

The equations  (${\cal C}$-polynomials) are also complicated,
somewhat unexpectedly for such a simple knot:

\begin{itemize}
\item[\underline{$n$-evolution:}]
\be\boxed{
\sum_{i=0}^5 Pol_i(q,A^2)F^{9_{46}}_{n+i}(A)=0}
\label{pol946}
\ee
where the polynomials $Pol_i(q,A^2)$ are of the degree at most 6 in $A^2$.
However, in this case, they are remarkably complicated,
explicit formulas being given in Appendix B.

\item[\underline{$A$-evolution:}]
\be
F^{9_{46}}_{n}(A)=\left(A^4q^{4n+2} + {[5]\over q^{2n+4}}\right)F^{9_{46}}_{n}(Aq)
+\left(A^4 \Big(\frac{[2]}{q}-[3]q^{2n}\Big)- {[4][5]\over [2]q^{4n+8}}\right)F^{9_{46}}_{n}(Aq^2)+
\nn\\
+\left(-A^{10}q^{8n+10}
+[3]A^4 \Big(\frac{1}{q^2}-\frac{[2]}{q^{2n+3}}\Big)
     +  {[4][5]\over [2]}{1\over q^{6n+12}} \right)F^{9_{46}}_{n}(Aq^3)+
\nn\\
+\left( A^{10}q^{4n+10}-A^8q^6+A^4\Big(\frac{[3][2]}{q^{4n+5}}-\frac{1}{q^{2n+4}}\Big)
-{[5]\over q^{8n+16}} \right)F^{9_{46}}_{n}(Aq^4)
+\nn\\
+ \left(\frac{A^8}{q^{2n-6}} - \frac{[2]A^4}{q^{6n+7}} + \frac{1}{q^{10n+20}}\right)
F^{9_{46}}_{n}(Aq^5)
\label{F935nconst}
\ee
or
\be\boxed{\begin{array}{c}
\sum_{j=0}{(-1)^{j+1}\over q^{2jn}}\left({{\rm Bin}_q(5,j)\over q^{4j}}+{A^4q^{6n+4}\over q^{2j}}{\rm Bin}_q(3,j-1)
-{A^4q^{4n+3}[2]\over q^{2j}}{\rm Bin}_q(3,j-2)
-\right.\cr
\cr
-\left.{A^{10}q^{10+20n}\over q^{-2jn}}{\rm Bin}_q(1,j-3)+A^8q^{8n+6}{\rm Bin}_q(1,j-4)
\right)F^{9_{46}}_{n}(Aq^j)=0
\end{array}}
\ee
Note that, for this knot similarly to the twist knots, the order of the both $n$-evolution and $A$-evolution equations is the same, though the equation itself is much simpler for the $A$-evolution.

\item[\underline{Non-local eq.:}]
The sum rule (\ref{sumrule1}) in this case is just the same as (\ref{sumruletwist2})
because $F_1^{9_{46}} = F_1^{5_1} = A^2(1+A^2)$ and $m_1=1, \ m_2=2$:
\be
\sum_{j=0}^{n+1} \frac{(-)^j}{q^{3jn }}\cdot\frac{[n+1]!}{[j]![n+1-j]!}
\cdot F^{9_{46}}_{n}(q^{j}\!A) = 0
\label{F946constn}
\ee

\item[\underline{Semi-local eq.:}]
\be\boxed{
F^{9_{46}}_{n+1}(A) + \frac{A^2(q^{2n}A^2+1)}{q^{2n}}F^{9_{46}}_{n}(q^2A)
- \sum_{j=0}^{\frac{n-1}{2}} (q^2-1)^{2j+1}\frac{A^{4(j+1)}q^{2j^2+(2j+3)n-3(j+1)}[n]!}{[n-2j-1]!}
F^{9_{46}}_{n-2j-1}(Aq^{2j+2})}
\ee
\end{itemize}

As usual, there are a lot of mixed evolution equations, both local equations and non-local ones\footnote{An example of an unusual non-local equation is
\be
F^{9_{46}}_{n+1}(A) = \sum_{j=0}^n (-)^jA^{2j+2}q^{5jn}\frac{[n]!}{[n-j]!}\{q\}^j
\Big(a_{j}\cdot q^{8n}A^2+b_j\cdot q^{4n} \Big)\cdot F^{9_{46}}_{n-j}(A)
\ee
with
\be
a_{0} = 1, \ \ \ \  a_{1} = q^{-6}, \ \ \ \ a_{2} = q^{-19}[3], \ \ \ \
a_{3} = q^{-37}[3]^2, \ \ \ \ a_{4} = q^{-60}([3]^2+q^{-6}[2]^2+q^{-10}), \ \ \ \
\ldots
\nn \\
b_{0} = 1, \ \ \ \  b_{1} = q^{-7}[2], \ \ \ \ b_{2} = q^{-19}[2]^2, \ \ \ \
b_{3} = q^{-37}([3][2]^2+q^{-6}), \ \ \ \
\ldots
\ee
These are  quantizations of sequences A063020 and A007858 from \cite{seqs}.
In particular, $y=\sum_{n=0}^\infty a_nx^{n+1}$ at $q=1$ is a solution of the equation
$y(y-1)(y^2-1)=x$.}. For instance, there are two rather simple local equations
\be
\!\!\!\!\!\!\!\!\!\!
F^{9_{46}}_{n+1}(A) =
A^4q^{4n}\cdot F^{9_{46}}_{n}(Aq) + \frac{A^2}{q^{4n}}\cdot F^{9_{46}}_{n}(Aq^2)
-  \{q^n\}\{q^{n-1}\}A^6\left( q^{2n+3}\cdot F^{9_{46}}_{n-2}(Aq^3)
+\frac{[2] }{q^{2n-8}}\cdot F^{9_{46}}_{n-2}(Aq^4)\right) +
\nn
\ee
\vspace{-0.7cm}
\be
+A^{10}q^{20}\{q^n\}\{q^{n-1}\}\{q^{n-2}\}\{q^{n-3}\}\cdot F^{9_{46}}_{n-4}(Aq^6)
\ee
and
\be
F^{9_{46}}_{n}(A) = q^{-2n}F^{9_{46}}_{n}(Aq) -  (q^{2n}-1)A^4\cdot F^{9_{46}}_{n-1}(Aq^2)
\label{simple946}
\ee
Note a spectacular simplicity of (\ref{simple946}),
especially impressive given the relative complexity of all other quantum ${\cal C}$-polynomials.
In variance with those, this equation has just three terms, still it is rather informative:
(\ref{simple946}) does {\it not} determine recursively only the lowest-degree item $A^{2n}$ in each
$F_n^{9_{46}} = A^{4n}q^{4n(n+1)} + \ldots + c_nA^{2n}$.

\section{Conclusion}

In this paper, we revisited the forgotten subject of quantum $C$-polynomials \cite{Gar05}
taking advantage of the new knowledge about the differential expansion of the HOMFLY-PT polynomials.
Most important, it allows us to lift our consideration from the Jones polynomials to the level of HOMFLY-PT polynomials.
Though we keep ourselves restricted to the symmetric representations $[n]$ only,
this extension allows us to complement the difference equations in $n$ by new
equations in $A=q^N$ (or in $N$, if one prefers). 
This promotes a set of $C$-polynomials to the entire non-commutative ring,
where the $n$-evolution and the $A$-evolution are treated on equal footing and serve
as the ring generators.
Similar structures should exist for ${\cal A}$-polynomials, but it is much more
difficult to study.

The differential expansion and thus the $C$-polynomials depend on the {\it defect} of the knot
\cite{Konodef}, so far we concentrated on the simplest case of defect zero.
All the smallest knots with this property (i.e. with the Alexander polynomial of degree one) belong to
the families of twisted knots and of double braids (much similar in properties to the twist knots),
the first one beyond these families is knot $9_{46}$.
We make use of an explicit knowledge of the symmetric DE coefficients for the twist family \cite{evo}
(see \cite{BiM} for review and references on the progress since then)
to give a rather general description of the $C$-polynomials in this case.
We consider also the example of $9_{46}$ to illustrate the relevance of our studies
beyond the twist family. For the knots discussed in this paper, it turns out that the difference equations in $n$ and in $A$ are of the same order (of order $k$ for the $k$-twist knots and of order 5 for knot $9_{46}$). It would interesting to establish if this is a universal feature of knot polynomials.

Surprisingly or not, the quantum ${\cal C}$-polynomials appear to be rather sophisticated.
Despite our strong belief that they are more fundamental than the ${\cal A}$-polynomials,
the theory does not yet look elementary.
Still it is much more straightforward and explicit.
However, new insights are still needed.
In particular, we did not truly exploit the powerful KNTZ-matrix description of the DE \cite{KNTZ},
which can be directly related to the ring structure of $C$-polynomials.
This is one of the many questions for further investigation.

\section*{Acknowledgements}

Our work is supported in part by the grant of the
Foundation for the Advancement of Theoretical Physics ``BASIS",
by  RFBR grants 19-01-00680 (A.Mir.) and 19-02-00815 (A.Mor.), 
by joint grants 19-51-50008-YaF-a (A.Mir.), 19-51-53014-GFEN-a, 18-51-05015-Arm-a.
The work was also partly funded by RFBR and NSFB according
to the research project 19-51-18006.

\section*{Appendix A.  $A$-evolution for twist knots
\label{Aevotwist}}

\subsection{Examples at low $k$}

\begin{itemize}
\item[\fbox{$k=-2$}]
\be
A^{2}q^{2n}\cdot F_{n}^{{\rm twist}_{-2}}(A)
- \Big(A^2q^{2n}-1\Big)\cdot F_{n}^{{\rm twist}_{-2}}(Aq)
- q^{2n}\cdot F_{n}^{{\rm twist}_{-2}}(Aq^2) =0
\label{Ashiftsfirst}
\ee
\item[\fbox{$k=-3$}]
\be
\Big(q^{2n+2}A^2-1\Big)\cdot A^6q^{4n+2}\cdot F_{n}^{{\rm twist}_{-3}}(A)-\nn\\
-\Big(q^{2n}A^2-1\Big)\Big(A^2q^{2n+2}+q^{2n+2}-1\Big)
\cdot A^4q^{2n+2}\cdot F_{n}^{{\rm twist}_{-3}}(Aq)-\nn\\
-\Big(q^{2n+2}A^2-1\Big)\Big(1+q^{2n}A^2(q^{2n+2}-1)\Big)\cdot F_{n}^{{\rm twist}_{-3}}(Aq^2)+
\nn\\
+\Big(q^{2n}A^2-1\Big)\cdot q^{4n}\cdot F_{n}^{{\rm twist}_{-3}}(Aq^3) =0
\ee
\item[\fbox{$k=-4$}]
\be
\Big(q^{2n+4}A^2-1\Big)\cdot A^{12}q^{6n+10}\cdot F_{n}^{{\rm twist}_{-4}}(A)-\nn\\
-\Big(q^{2n+4}A^2-1\Big)\Big(q^{2n}A^2-1\Big)
 (A^2+1)\cdot A^8q^{4n+10}\cdot F_{n}^{{\rm twist}_{-4}}(Aq) - \nn \\
-\Big(q^{2n+2}A^2-1\Big)\left\{1+q^{2n}A^2\Big((q^2+1)q^{2n+2}-(q^4+1)\Big)+A^4q^{4n+4}\right\}
A^4q^{2n+8}\cdot F_{n}^{{\rm twist}_{-4}}(Aq^2) + \nn \\
+ \Big(q^{2n+4}A^2-1\Big)\Big(q^{2n}A^2-1\Big)
 (q^{4n+4}A^2+1)\cdot F_{n}^{{\rm twist}_{-4}}(Aq^3)+\nn\\
+  \Big(q^{2n}A^2-1\Big)\cdot q^{6n}\cdot F_{n}^{{\rm twist}_{-4}}(Aq^4)=0
\ee
\vspace{-4cm}
\item[\fbox{$k=-5$}]
\hspace{-1.5cm}
\parbox{17cm}
{\footnotesize
\vspace{5cm}
\be
\Big(A^2q^{2n+6} -1\Big)\Big(A^2q^{2n+4} -1\Big)
\cdot A^{20}q^{8n+26}\cdot F_{n}^{{\rm twist}_{-5}}(A) -\nn \\
-\Big(A^2q^{2n+6} -1\Big)\Big(A^2q^{2n} -1\Big)\Big(A^4q^{2n+4}+(A^2+1)(q^{2n+4}-1) \Big)
\cdot A^{16}q^{6n+26}
\cdot F_{n}^{{\rm twist}_{-5}}(Aq) -\nn \\
- \Big(A^2q^{2n+6} -1\Big)\Big(A^2q^{2n+2} -1\Big)
\Big((q^{2n+4}-1)A^4 q^{2n+2}+(q^{2n+4}-1)(q^{2n}+q^{2n-2}-1)A^2q^2+(q^2+1)\Big)
\cdot A^{10}q^{4n+22}\cdot F_{n}^{{\rm twist}_{-5}}(Aq^2) + \nn \\
+ \Big(A^2q^{2n+4} -1\Big)\Big(A^2q^{2n} -1\Big)
\Big((q^2+1)A^4 q^{6n+10}+(q^{2n+4}-1)(q^{2n}+q^{2n-2}-1)A^2q^{2n+6}+(q^{2n+4}-1)\Big)
\cdot A^{6}q^{2n+14} \cdot F_{n}^{{\rm twist}_{-5}}(Aq^3) + \nn\\
+ \Big(A^2q^{2n+6} -1\Big)\Big(A^2q^{2n} -1\Big)
\Big((q^{2n+4}-1)A^4 q^{6n+8}+(q^{2n+4}-1)A^2q^{2n+2}+1\Big) \cdot F_{n}^{{\rm twist}_{-5}}(Aq^4) -\nn\\
- \Big(A^2q^{2n+2} -1\Big)\Big(A^2q^{2n} -1\Big)\cdot q^{8n}\cdot F_{n}^{{\rm twist}_{-5}}(Aq^5)
=0
\label{Ashiftslast}
\ee
}
\end{itemize}

\subsection{General answer}

The general structure of the answer is
\be
 \sum_{j=0}^{k-1} \epsilon_j G_{n,j}^{(k)}
\cdot F_{n}^{{\rm twist}_{-k}}(Aq^j)
+\epsilon_k\cdot q^{2(k-1)n} \cdot \prod_{i=0}^{\frac{k-3}{2}} \Big(A^2q^{2n+2i}-1\Big)
\cdot F_{n}^{{\rm twist}_{-k}}(Aq^k) = 0
\ee
\be
G^{(k)}_{n,j} = &\Big(A^2q^{2n+2j-2}-1\Big)\cdot\prod_{i=0}^{\ent\left(\frac{j-3}{2}\right)}\Big(A^2q^{2n+2i}-1\Big) \cdot \prod_{i=\ent\left(\frac{k+j}{2}\right)}^{k-2}\Big(A^2q^{2n+2i}-1\Big)\times\nn\\ \nn\\
& \times A^{k^2-(j+1)k+4\fr\left(\frac{k}{2}\right)\fr\left(\frac{j}{2}\right)}q^{2(k-j-1)n+\sigma_{k,j}}\cdot g^{(k)}_{n,j}
\ee
where $\epsilon_j = \sqrt{2}\cdot\cos\left(\frac{(2j+1)\pi}{4}\right)$, and we conjecture that
\be
\sigma_{k,j}=\ent\left(\frac{k(k-2)(2k-3)}{4}\right)-{j(j-1)\over 2}k+\ent\left(\frac{j}{2}\right)^2+\ent\left(\frac{j}{2}\right)
-2(-1)^j\fr\left(\frac{k}{2}\right)\ent\left(\frac{j}{2}\right)
\ee
and
\be
g^{(k)}_{n,0} &=& 1
\ee
\be
g^{(k)}_{n,1} &=& \left\{\begin{array}{ccl}
k \ {\rm even} && \xi_k(A) \\ \\
k \ {\rm odd} &&q^{2n+k-1}\cdot\xi_{k+1}(A) - \xi_{k-1}(A)
\end{array}\right.
\ee
\be
g^{(k)}_{n,2} &=& \left\{\begin{array}{ccl}
k \ {\rm even} && \displaystyle{A^2q^{4n}\sum_{i=1}^{k\over 2}q^{k-4+2i}\xi_{2i}(A)
+\Big(1-A^2q^{2n}(1+q^k)\Big)\sum_{i=1}^{{k\over 2}-1}q^{2(i-1)}\xi_{2i}(A)} \\ \\
k \ {\rm odd} &&
\displaystyle{A^2q^{4n}\sum_{i=1}^{k-1\over 2}q^{k-3+2i}\xi_{2i}(A)
+\Big(1-A^2q^{2n}\Big)\sum_{i=1}^{k-1\over 2}q^{2(i-1)}\xi_{2i}(A)
+\sum_{i=0}^{k-3\over 2}q^{2i+4}\xi_{2i}(A)}
\end{array}\right.
\ee
\be
\ldots
\ee
where $\xi_l(A):=\displaystyle{\frac{A^l-1}{A^2-1}}$ is a polynomial of $A$ only (since $j$ is even in the formulas above). Similarly, expressions for $g^{(k)}_{n,j}$ at $j>2$ involve $\displaystyle{\frac{(q^mA)^l-1}{(q^mA)^2-1}}$ with $m\le j-2$.

\subsection{Examples of $G^{(k)}_{n,j}$}

\be
G^{(k)}_{n,0} = & \prod_{i=\ent\left(\frac{k}{2}\right)}^{k-2}\Big(A^2q^{2n+2i}-1\Big)
\cdot A^{k(k-1)}q^{2(k-1)n+\sigma_k}
\ee

\be
G^{(2)}_{n,1} =& \Big(A^2q^{2n}-1\Big)& \nn \\
G^{(3)}_{n,1} =& \Big(A^2q^{2n}-1\Big)&\cdot\Big(q^{2n+2}(A^2+1)-1\Big)\cdot A^4q^{2n+2} \nn \\
G^{(4)}_{n,1} =& \Big(A^2q^{2n}-1\Big)&\cdot\Big(A^2q^{2n+4}-1\Big)(A^2+1)\cdot A^8q^{4n+10} \nn \\
G^{(5)}_{n,1} =& \Big(A^2q^{2n}-1\Big)&\cdot\Big(A^2q^{2n+6}-1\Big)\cdot
\Big(q^{2n+4}(A^4+A^2+1)-(A^2+1) \Big)  \cdot A^{16}q^{6n+26} \nn \\
G^{(6)}_{n,1} =& \Big(A^2q^{2n}-1\Big)&\cdot\Big(A^2q^{2n+6}-1\Big)\Big(A^2q^{2n+8}-1\Big)\cdot
\Big(A^4 +A^2+1\Big)  \cdot A^{24}q^{8n+54} \nn \\
G^{(7)}_{n,1} =& \Big(A^2q^{2n}-1\Big)&\cdot\Big(A^2q^{2n+8}-1\Big)\Big(A^2q^{2n+10}-1\Big)\cdot
\Big(q^{2n+6}(A^6+A^4 +A^2+1) - (A^4+A^2+1)\Big)  \cdot A^{36}q^{10n+96} \nn \\
\ldots
\ee

{\footnotesize
\be
\!\!\!\!\!\!\!\!\!\!\!\!\!\!\!\!\!\!
 G^{(3)}_{n,2} =& \Big(A^2q^{2n+2}-1\Big)&\!\!\!\!\Big(A^2q^{2n}(q^{2n+2}-1)+1\Big)
 \nn \\
 \!\!\!\!\!\!\!\!\!\!\!\!\!\!\!\!\!\!
G^{(4)}_{n,2} =& \Big(A^2q^{2n+2}-1\Big)&
\!\!\!\!\left\{A^4q^{4n+4}+\Big(q^{2n+2}(q^2+1)-(q^4+1)\Big)A^2q^{2n}+1\right\}
\cdot A^4q^{2n+8} \nn \\
\!\!\!\!\!\!\!\!\!\!\!\!\!\!\!\!\!\!
 G^{(5)}_{n,2} =& \Big(A^2q^{2n+2}-1\Big)&\!\!\!\!\Big(A^2q^{2n+6}-1\Big)
\Big(A^4q^{2n+2}(q^{2n+4}-1)+(q^{2n+4}-1)(q^{2n+2}+q^{2n }-q^2)\cdot A^2  + (q^2+1)\Big)
\cdot A^{10}q^{4n+22} \nn \\
\!\!\!\!\!\!\!\!\!\!\!\!\!\!\!\!\!\!
G^{(6)}_{n,2} =& \Big(A^2q^{2n+2}-1\Big)&\!\!\!\! \Big(A^2q^{2n+8}-1\Big) \cdot
\Big\{A^6q^{4n+8} +  (q^{4n+8}+q^{4n+6}-q^{2n+8}-q^{2n+2})A^4  + \nn \\
 &&
+\Big((q^4+q^2+1)q^{4n+4}-(q^4+q^2+q^{-2}+q^{-4})q^{2n+4}+q^2\Big)A^2+(q^2+1)\Big\}
\cdot A^{18}q^{6n+50} \nn \\
\!\!\!\!\!\!\!\!\!\!\!\!\!\!\!\!\!\!
G^{(7)}_{n,2} =& \Big(A^2q^{2n+2}-1\Big)&\!\!\!\! \Big(A^2q^{2n+8}-1\Big)\Big(A^2q^{2n+10}-1\Big) \cdot
\Big\{(q^{2n+6}-1)\Big(A^6q^{2n+4}+\Big((q^2+1)q^{2n+2}-q^{4}\Big)A^4
+ \Big((q^4+q^2+1)q^{2n }-(q^4+q^2)  \Big)A^2\Big)+ \nn \\
 &&
+(q^4+q^2+1)  \Big\}
\cdot A^{28}q^{8n+90} \nn \\
\ldots
\ee
}

{\footnotesize
\be
\!\!\!\!\!\!\!\!\!\!\!\!\!\!\!\!\!\!
G^{(4)}_{n,3} =& \Big(A^2q^{2n}-1\Big)\Big(A^2q^{2n+4}-1\Big)
& \Big(A^2q^{4n+4}+1\Big) \nn \\
\!\!\!\!\!\!\!\!\!\!\!\!\!\!\!\!\!\!
G^{(5)}_{n,3} =& \Big(A^2q^{2n}-1\Big)\Big(A^2q^{2n+4}-1\Big)
& \Big(A^4q^{6n+10}(q^2+1)+(q^{2n+4}-1)(q^{2n}+q^{2n-2}-1)\cdot A^2q^{2n+6} + (q^{2n+4}-1)\Big)
\cdot A^{6}q^{2n+14} \nn \\
\!\!\!\!\!\!\!\!\!\!\!\!\!\!\!\!\!\!
G^{(6)}_{n,3} =& \Big(A^2q^{2n}-1\Big)\Big(A^2q^{2n+4}-1\Big)& \Big(A^2q^{2n+8}-1\Big)\cdot
(q^2+1)\cdot \Big(A^4q^{4n+8} +(q^{4n+4}-q^{2n+4}+q^{2n+2}-q^{2n}+1)A^2q^2+1\Big)  \cdot A^{12}q^{4n+38}
\nn \\
\!\!\!\!\!\!\!\!\!\!\!\!\!\!\!\!\!\!
G^{(7)}_{n,3} =& \Big(A^2q^{2n}-1\Big)\Big(A^2q^{2n+4}-1\Big)& \Big(A^2q^{2n+10}-1\Big)\cdot
 \Big\{(q^4+q^2+1)A^6q^{6n+16} +(q^{2n+6}-1)\Big((q^2+1)\Big((q^2+1)q^{4n}-q^{2n+2}\Big)A^4q^{8} + \nn\\
&& +\Big((q^4+q^2+1)q^{4n+4}-(q^2+1)q^{2n+6}-q^{2n}+q^2+1\Big)A^2q^2
 +q^2+1\Big)\Big\}  \cdot A^{22}q^{6n+78}
\nn \\
\ldots
\ee
}

{\footnotesize
\be
\!\!\!\!\!\!\!\!\!\!\!\!\!\!\!\!\!\!\!\!\!\!\!\!\!\!\!\!\!\!\!\!\!\!\!\!\!\!\!\!\!\!\!\!
G^{(5)}_{n,4} =& \Big(A^2q^{2n}-1\Big)\Big(A^2q^{2n+6}-1\Big)&\!\!\!\!\!\!
\Big((A^4q^{4n+12}+A^2q^{4n+6}+1)   -(A^2q^{4n+6} +1)\cdot  A^2q^{2n+2}  \Big)
\nn \\
\!\!\!\!\!\!\!\!\!\!\!\!\!\!\!\!\!\!\!\!\!\!\!\!\!\!\!\!\!\!\!\!\!\!\!\!\!\!\!\!\!\!\!\!
G^{(6)}_{n,4} =& \Big(A^2q^{2n}-1\Big) \Big(A^2q^{2n+6}-1\Big)&\!\!\!\!\!\!
\Big\{ (q^2+1)A^6q^{8n+16}+\Big((q^4+q^2+1)q^{4n+2}-(q^4+q^2+q^{-2}+q^{-4})q^{2n+2}+1\Big)A^4q^{4n+10}
+ \nn \\
&& + \Big((q^2+1)q^{4n+6}-(q^6+1)q^{2n+2}\Big)A^2 + 1
\Big\}
  \cdot A^{6}q^{2n+24} \nn \\
\!\!\!\!\!\!\!\!\!\!\!\!\!\!\!\!\!\!\!\!\!\!\!\!\!\!\!\!\!\!\!\!\!\!\!\!\!\!\!\!\!\!\!\!
G^{(7)}_{n,4} =& \Big(A^2q^{2n}-1\Big)\Big(A^2q^{2n+6}-1\Big)& \!\!\!\!\!\!\Big(A^2q^{2n+10}-1\Big)\cdot
 \Big\{(q^{2n+6}-1)\Big((q^2+1)A^6q^{6n+14} +\nn \\
  && \!\!\!\!\!\!\!\!\!\!\!\!\!\!\!\!\!\!\!\!\!\!\!\!\!\!\!\!\!\!\!\!\!\!\!\!\!\!\!\!\!\!\!\!\!\!\!\!\!\!
  \!\!\!\!\!\!\!\!\!\!\!\!\!\!\!\!\!\!\!\!\!\!\!\!\!\!\!\!\!\!\!\!\!\!\!\!\!\!\!\!\!\!\!\!\!\!\!\!\!\!
  +\Big((q^4+q^2+1)q^{6n+10}-(q^2+1)q^{4n+12} -q^{4n+6 } + (q^2+1)q^{2n+6 }\Big)A^4
  +(q^2+1) \Big((q^2+1)q^{2n+2}-q^{4}\Big)A^2  \Big)
 +(q^4+q^2+1) \Big\}  \cdot A^{14}q^{4n+58}
\nn \\
\ldots
\ee
}

{\footnotesize
\be
G^{(6)}_{n,5} =& \Big(A^2q^{2n}-1\Big) \Big(A^2q^{2n+2}-1\Big) \Big(A^2q^{2n+8}-1\Big)&
 \Big(A^4q^{8n+16} + A^2q^{4n+8}+1\Big) \nn  \\
G^{(7)}_{n,5} =& \Big(A^2q^{2n}-1\Big) \Big(A^2q^{2n+2}-1\Big) \Big(A^2q^{2n+8}-1\Big)&
 \Big\{(q^4+q^2+1)A^6q^{10n+26}+\Big((q^4+q^2+1)\cdot q^{8n+16}-(q^2+1)\cdot q^{6n+18}\Big)\cdot A^4
 + \nn \\
&&+\Big((q^2+1)\cdot q^{4n+8}-q^{2n+10}\Big)\cdot A^2 + (q^{2n+6}-1)\Big\}\cdot A^8q^{2n+34}  \nn \\
\ldots
\ee
}

{\footnotesize
\be
G^{(7)}_{n,6} =& \Big(A^2q^{2n}-1\Big) \Big(A^2q^{2n+2}-1\Big) \Big(A^2q^{2n+10}-1\Big)&
 \Big\{\Big(A^6q^{8n+20}+A^4q^{4n+10}+A^2\Big)(q^{2n+6}-1 )q^{2n+4}\  +\ 1\Big\} \nn  \\
\ldots
\ee
}

\section*{Appendix B. Details on knot $9_{46}$
\label{ap946}}

Explicit expressions for the entries of (\ref{pol946}):

{\footnotesize
$$
\{q^{n+2}\}q^{7n+12}A^6\Bigg([2]^4[3]^2\{q\}^2\alpha^{(1)}_{0,1}-q^{2n-1}[2]^3\alpha^{(1)}_{0,2}
-q^{4n+4}[2]^2\alpha^{(1)}_{0,3}-q^{6n+14}[4][2]\alpha^{(1)}_{0,4}+
$$
$$
+A^2\Big[q^{2n+13}[6][4][3]^2[2]^4\{q\}^3-q^{4n}[2]^4\alpha^{(1)}_{1,1}+q^{6n+3}[2]\alpha^{(1)}_{1,2}
-q^{8n+14}[3]\alpha^{(1)}_{1,3}+q^{10n+33}[2]^3[4]^3(q^8[2]-[4])\Big]\Bigg)F^{9_{46}}_{n+1}(A)-
$$
$$
-A^4q^{4n+6}\Bigg(q[3][2]^5\{q\}^2\alpha^{(2)}_{0,1}+q^{2n}[2]^2\alpha^{(2)}_{0,2}
+q^{4n+4}\alpha^{(2)}_{0,3}+q^{6n+15}[4]\alpha^{(2)}_{0,4}
+q^{8n+38}[4]^4[2]^4-
$$
$$
-A^2\Big[q^{2n}[2]^4[3]\{q\}^2\alpha^{(2)}_{1,1}-q^{4n}[2]^2\alpha^{(2)}_{1,2}+q^{6n+4}\alpha^{(2)}_{1,3}
+q^{8n+16}[2]^2\alpha^{(2)}_{1,4}+q^{10n+28}[4]^2[2]^2\alpha^{(2)}_{1,5}\Big]-
$$
$$
-A^4\Big[q^{4n+16}[6][4][3]^3[2]^3\{q\}^3
-q^{6n+5}[3][2]^3\alpha^{(2)}_{2,1}-q^{8n+6}\alpha^{(2)}_{2,2}-q^{10n+13}[2]\alpha^{(2)}_{2,3}
-q^{12n+26}[4]^2\alpha^{(2)}_{2,4}\Big]-
$$
$$
-A^6\Big[q^{8n+24}[6]^2[3][2]^5\{q\}^3-q^{10n+12}[2]^4\alpha^{(2)}_{3,1}+q^{12n+16}[2]^2\alpha^{(2)}_{3,2}
+q^{14n+22}\alpha^{(2)}_{3,3}+q^{16n+45}[2]^3[4]^3(q^8[4]-[2])
\Big]\Bigg)F^{9_{46}}_{n+2}(A)+
$$
$$
+A^2q^{2n+5}\Bigg(q^{-1}[3][2]^4\{q\}^2\alpha^{(3)}_{0,1}-q^{2n}[2]^3\alpha^{(3)}_{0,2}-q^{4n+6}[2]\alpha^{(3)}_{0,3}
-q^{6n+29}[2]^3[4]^3([3]+q^{10})+
$$
$$
+A^2\Big[q^{2n+6}[3]^2[2]^4\{q\}^3\alpha^{(3)}_{1,1}+q^{4n+3}[2]^2\alpha^{(3)}_{1,2}+q^{6n+9}\alpha^{(3)}_{1,3}
+q^{8n+26}[4]^2[2]\alpha^{(3)}_{1,4}\Big]-
$$
$$
-A^4\Big[q^{4n+1}[3][2]^4\{q\}^2\alpha^{(3)}_{2,1}-q^{6n+3}[2]^4\alpha^{(3)}_{2,2}+q^{8n+9}[2]^2\alpha^{(3)}_{2,3}
+q^{10n+20}[4]\alpha^{(3)}_{2,4}+q^{12n+41}[4]^4[2]^4\Big]-
$$
$$
-A^6\Big[q^{6n+16}[6][4][3]^2[2]^4\{q\}^3+q^{8n+9}[2]^4\alpha^{(3)}_{3,1}-q^{10n+10}[2]\alpha^{(3)}_{3,2}
-q^{12n+17}\alpha^{(3)}_{3,3}-q^{14n+39}[4]^3[3][2]\alpha^{(3)}_{3,4}\Big]-
$$
$$
-A^8\Big[q^{10n+28}[6][3]^2[2]^5\{q\}^3+q^{12n+21}[2]^4\alpha^{(3)}_{4,1}-q^{14n+28}[2]\alpha^{(3)}_{4,2}
-q^{16n+37}[4][2]\alpha^{(3)}_{4,3}\Big]-
$$
$$
-A^{10}\Big[q^{14n+39}[6][3]^2[2]^4\{q\}^3-q^{16n+32}[2]^3\alpha^{(3)}_{5,1}-q^{18n+39}\alpha^{(3)}_{5,2}
-q^{20n+60}[4]^3[2]^4
\Big]\Bigg)F^{9_{46}}_{n+3}(A)-
$$
$$
-\Bigg(q^8[6][4][3]^2[2]^3\{q\}^3+q^{2n+1}[3][2]^2\{q\}\alpha^{(4)}_{0,1}-q^{4n+9}[4][2]\{q\}\alpha^{(4)}_{0,2}+
$$
$$
+A^2\Big[q^{2n-2}[3][2]^4\{q\}^2\alpha^{(4)}_{1,1}-q^{4n-2}[2]^2\alpha^{(4)}_{1,2}-q^{6n+6}\alpha^{(4)}_{1,3}
-q^{8n+27}[4]^3[2]^4\Big]-
$$
$$
-A^4\Big[q^{4n+3}[3][2]^5\{q\}^2\alpha^{(2)}_{0,1}-q^{6n+2}[2]^2\alpha^{(2)}_{0,2}-q^{8n+8}\alpha^{(2)}_{0,3}
-q^{10n+31}[4]^3[2]^2\alpha^{(4)}_{2,1}\Big]-
$$
$$
-A^6\Big[q^{6n+10}[3]^2[2]^4\{q\}^2\alpha^{(4)}_{3,1}-q^{8n+8}[2]^2\alpha^{(4)}_{3,2}-q^{10n+14}\alpha^{(4)}_{3,3}
-q^{12n+32}[4]^2[2]^2\alpha^{(4)}_{3,4}\Big]-
$$
$$
-A^8\Big[q^{8n+20}[6][3]^2[2]^4\{q\}^3+q^{10n+13}[2]^3\alpha^{(4)}_{4,1}+q^{12n+20}\alpha^{(4)}_{4,2}
+q^{14n+37}[4]^2[2]^3\alpha^{(4)}_{4,3}\Big]
\Big]\Bigg)F^{9_{46}}_{n+4}(A)+
$$
$$
\hspace{-2cm}
-q^8\Big([3][2]^4\{q\}^2\beta_1+q^{2n}[2]^2\beta_2+q^{4n+8}\beta_3+q^{6n+27}[4]^3[2]^4\Big)\Big(F^{9_{46}}_{n+3}(A)
-q^{2n+4}A^2F^{9_{46}}_{n+2}(A)+\{q^{n+2}\}q^{5n+6}A^4F^{9_{46}}_{n+1}(A)+\{q^{n+1}\}\{q^{n+2}\}q^{8n+7}A^6
F^{9_{46}}_{n}(A)\Big)+\nn
$$
$$
\hspace{-2cm}
+q^{10n+22}\Bigg(q^5[6][2]^4[3]^2\{q\}^3-q^{2n}[2]^3\beta_4-q^{4n+1}\beta_5-q^{6n+25}[8][4]^2[2]^3\Bigg)
\Bigg(q^{6n+7}A^{12}F^{9_{46}}_{n+2}(A)+q^{3n+2}[2]\{q^{n+2}\}A^{10}F^{9_{46}}_{n+1}(A)
+\{q^{n+1}\}\{q^{n+2}\}A^8F^{9_{46}}_{n}(A)\Bigg)-
$$
$$
\hspace{-2cm}
-\{q\}[2]^2\Bigg([3][2]^2\{q\}\beta_6+q^{2n}[2]\beta_7+q^{4n+14}[3][4]^2(q^7[3]-[2])\Bigg)
\Bigg(\Big[q^{6n+14}A^6+q^{4n+8}A^4-q^{2n}A^2-q^{-6}\Big]F^{9_{46}}_{n+5}(A)-A^{10}F^{9_{46}}_{n+4}(A)\Bigg)=0
$$
}

\bigskip

\noindent
where the coefficients involve the following structures:
{\footnotesize
\be
\alpha^{(1)}_{0,1}=q^{{20}}-2q^{14}-q^{12}-2q^{10}-3q^{8}-4q^{6}-3q^{4}-q^{2}-1\nn\\
\alpha^{(1)}_{0,2}=q^{{36}}+4q^{{34}}+8q^{{32}}+11q^{30}+12q^{28}+16q^{26}+28q^{24}+46q^{22}+64q^{20}+74q^{18}+\nn\\
+82q^{16}+76q^{14}+60q^{12}+31q^{10}+11q^{8}-q^{6}-5q^{4}-4q^{2}-2\nn\\
\alpha^{(1)}_{0,3}=-2q^{38}-8q^{36}-19q^{34}-36q^{32}-60q^{30}-94q^{28}-139q^{26}-192q^{24}-242q^{22}-278q^{20}-\nn\\
-287q^{18}-263q^{16}-207q^{14}-137q^{12}-73q^{10}-29q^{8}-3q^{6}+5q^{4}+4q^{2}+1\nn\\
\alpha^{(1)}_{0,4}=q^{32}+4q^{30}+8q^{28}+14q^{26}+24q^{24}+40q^{22}+57q^{20}+70q^{18}+74q^{16}+71q^{14}+\nn\\
+61q^{12}+47q^{10}+30q^{8}+13q^{6}+q^{4}-2q^{2}-1\nn\\
\alpha^{(1)}_{1,1}=q^{36}+5q^{34}+8q^{32}+16q^{30}+15q^{28}+23q^{26}+18q^{24}+27q^{22}+25q^{20}+33q^{18}+\nn\\
+29q^{16}+26q^{14}+16q^{12}+8q^{10}+q^{8}+2q^{4}+2q^{2}+1\nn\\
\alpha^{(1)}_{1,2}=2q^{42}+14q^{40}+45q^{38}+108q^{36}+202q^{34}+313q^{32}+409q^{30}+469q^{28}+486q^{26}+\nn\\
+467q^{24}+414q^{22}+333q^{20}+228q^{18}+116q^{16}+24q^{14}-23q^{12}-25q^{10}-4q^{8}+11q^{6}+11q^{4}+5q^{2}+1\nn\\
\alpha^{(1)}_{1,3}=q^{36}+7q^{34}+23q^{32}+59q^{30}+117q^{28}+183q^{26}+231q^{24}+240q^{22}+201q^{20}+126q^{18}+\nn\\
+35q^{16}-52q^{14}-113q^{12}-134q^{10}-117q^{8}-77q^{6}-37q^{4}-12q^{2}-2\nn\\
\alpha^{(2)}_{0,1}=q^{24}+2q^{20}+3q^{16}-q^{14}+3q^{12}+q^{10}+4q^{8}+2q^{4}+1\nn\\
\alpha^{(2)}_{0,2}=q^{40}+5q^{38}+10q^{36}+15q^{34}+16q^{32}+13q^{30}-2q^{28}-23q^{26}-39q^{24}-42q^{22}-46q^{20}-\nn\\
-57q^{18}-61q^{16}-47q^{14}-24q^{12}-5q^{10}+4q^{8}+9q^{6}+9q^{4}+6q^{2}+2\nn
\\
\alpha^{(2)}_{0,3}=-2q^{44}-12q^{42}-34q^{40}-66q^{38}-93q^{36}-85q^{34}+187q^{30}+454q^{28}+730q^{26}+936q^{24}+\nn\\
+1025q^{22}+992q^{20}+856q^{18}+650q^{16}+417q^{14}+208q^{12}+62q^{10}-12q^{8}-30q^{6}-20q^{4}-7q^{2}-1\nn
\ee
\be
\alpha^{(2)}_{0,4}=q^{36}+5q^{34}+11q^{32}+15q^{30}+5q^{28}-35q^{26}-117q^{24}-229q^{22}-336q^{20}-398q^{18}-\nn\\
-397q^{16}-339q^{14}-248q^{12}-154q^{10}-77q^{8}-25q^{6}-q^{4}+3q^{2}+1\nn\\
\alpha^{(2)}_{1,1}=q^{28}+2q^{26}+4q^{24}+4q^{22}+3q^{20}+q^{18}-q^{16}-2q^{14}-8q^{12}-10q^{10}-11q^{8}-7q^{6}-5q^{4}-2q^{2}-1\nn\\
\alpha^{(2)}_{1,2}=q^{42}+7q^{40}+19q^{38}+37q^{36}+53q^{34}+64q^{32}+62q^{30}+63q^{28}+78q^{26}+114q^{24}+\nn\\
+149q^{22}+180q^{20}+204q^{18}+210q^{16}+177q^{14}+113q^{12}+44q^{10}-16q^{6}-14q^{4}-7q^{2}-2\nn\\
\alpha^{(2)}_{1,3}=3q^{{46}}+22q^{{44}}+78q^{{42}}+189q^{{40}}+359q^{{38}}+579q^{{36}}+821q^{{34}}+1053q^{{32}}
+1272q^{{30}}+1523q^{{28}}+1839q^{{26}}+\nn\\
+2169q^{{24}}+2391q^{{22}}+2399q^{{20}}+2153q^{{18}}+1696q^{{16}}+1134q^{{14}}+610q^{{12}}+240q^{{10}}+50q^{8}
-12q^{6}-15q^{4}-6q^{2}-1\nn\\
\alpha^{(2)}_{1,4}=-3q^{38}-15q^{36}-41q^{34}-85q^{32}-145q^{30}-219q^{28}-298q^{26}-382q^{24}-468q^{22}-562q^{20}-\nn\\
-643q^{18}-686q^{16}-660q^{14}-562q^{12}-414q^{10}-259q^{8}-135q^{6}-57q^{4}-17q^{2}-3\nn\\
\alpha^{(2)}_{1,5}=q^{26}+3q^{24}+5q^{22}+7q^{20}+8q^{18}+10q^{16}+13q^{14}+18q^{12}+20q^{10}+18q^{8}+13q^{6}+8q^{4}+3q^{2}+1\nn\\
\alpha^{(2)}_{2,1}=q^{34}+5q^{32}+11q^{30}+17q^{28}+19q^{26}+21q^{24}+20q^{22}+25q^{20}+27q^{18}+30q^{16}+\nn\\
+26q^{14}+24q^{12}+17q^{10}+9q^{8}+3q^{6}+1\nn\\
\alpha^{(2)}_{2,2}=-q^{{44}}-14q^{{42}}-69q^{{40}}-210q^{{38}}-464q^{{36}}-813q^{{34}}-1193q^{{32}}-1527q^{{30}}
-1754q^{{28}}-1853q^{{26}}-\nn\\
-1833q^{{24}}-1715q^{{22}}
-1509q^{{20}}-1229q^{{18}}-908q^{{16}}-601q^{{14}}-356q^{{12}}-194q^{{10}}-105q^{8}-59q^{6}-30q^{4}-11q^{2}-2\nn\\
\alpha^{(2)}_{2,3}=-q^{42}+5q^{40}+34q^{38}+112q^{36}+252q^{34}+447q^{32}+659q^{30}+836q^{28}+930q^{26}+925q^{24}+\nn\\
+834q^{22}+690q^{20}+525q^{18}+366q^{16}+236q^{14}+144q^{12}+86q^{10}+53q^{8}+33q^{6}+17q^{4}+6q^{2}+1\nn\\
\alpha^{(2)}_{2,4}=q^{30}-6q^{26}-19q^{24}-33q^{22}-44q^{20}-47q^{18}-41q^{16}-29q^{14}-18q^{12}-8q^{10}-2q^{8}-2q^{6}-4q^{4}-3q^{2}-1\nn\\
\alpha^{(2)}_{3,1}=q^{36}+5q^{34}+9q^{32}+13q^{30}+15q^{28}+21q^{26}+25q^{24}+25q^{22}+21q^{20}+23q^{18}+\nn\\
+28q^{16}+26q^{14}+20q^{12}+9q^{10}+9q^{8}+2q^{6}+3q^{4}+1\nn\\
\alpha^{(2)}_{3,2}=2q^{40}+13q^{38}+38q^{36}+82q^{34}+138q^{32}+200q^{30}+252q^{28}+285q^{26}+282q^{24}+\nn\\
+254q^{22}+214q^{20}+180q^{18}+141q^{16}+99q^{14}+57q^{12}+31q^{10}+19q^{8}+14q^{6}+9q^{4}+4q^{2}+1\nn\\
\alpha^{(2)}_{3,3}=-q^{42}-10q^{40}-43q^{38}-123q^{36}-270q^{34}-483q^{32}-723q^{30}-922q^{28}-1013q^{26}-959q^{24}-\nn\\
-774q^{22}-522q^{20}-277q^{18}-94q^{16}+6q^{14}+35q^{12}+26q^{10}+8q^{8}-4q^{6}-7q^{4}-4q^{2}-1\nn\\
\alpha^{(3)}_{0,1}=q^{24}+q^{22}+2q^{20}+q^{18}+q^{16}-q^{14}-2q^{12}-3q^{10}-4q^{8}-5q^{6}-4q^{4}-2q^{2}-1\nn\\
\alpha^{(3)}_{0,2}=q^{36}+4q^{34}+7q^{32}+10q^{30}+11q^{28}+13q^{26}+13q^{24}+15q^{22}+20q^{20}+27q^{18}+\nn\\
+34q^{16}+36q^{14}+35q^{12}+24q^{10}+12q^{8}+q^{6}-3q^{4}-3q^{2}-1\nn\\
\alpha^{(3)}_{0,3}=-2q^{38}-10q^{36}-26q^{34}-49q^{32}-75q^{30}-101q^{28}-127q^{26}-155q^{24}-184q^{22}-214q^{20}-\nn\\
-238q^{18}-248q^{16}-234q^{14}-190q^{12}-126q^{10}-64q^{8}-20q^{6}+3q^{2}+1\nn
\\
\alpha^{(3)}_{1,1}=q^{12}+q^{10}+q^{8}+q^{6}+3q^{4}+2q^{2}+1\nn\\
\alpha^{(3)}_{1,2}=q^{34}+2q^{32}+q^{30}-4q^{28}-6q^{26}-7q^{24}-9q^{22}-21q^{20}-31q^{18}-34q^{16}-34q^{14}-\nn\\
-37q^{12}-38q^{10}-29q^{8}-12q^{6}-q^{4}+2q^{2}+1\nn\\
\alpha^{(3)}_{1,3}=-2q^{36}-9q^{34}-21q^{32}-30q^{30}-25q^{28}+5q^{26}+65q^{24}+152q^{22}+248q^{20}+324q^{18}+\nn\\
+355q^{16}+334q^{14}+271q^{12}+189q^{10}+113q^{8}+58q^{6}+24q^{4}+7q^{2}+1\nn\\
\alpha^{(3)}_{1,4}=q^{18}+2q^{16}+3q^{14}+q^{12}-3q^{10}-9q^{8}-11q^{6}-9q^{4}-5q^{2}-2\nn\\
\alpha^{(3)}_{2,1}=q^{28}+2q^{26}+5q^{24}+6q^{22}+8q^{20}+7q^{18}+9q^{16}+7q^{14}+4q^{12}-2q^{10}-4q^{8}-4q^{6}-4q^{4}-2q^{2}-1\nn\\
\alpha^{(3)}_{2,2}=q^{38}+4q^{36}+8q^{34}+14q^{32}+14q^{30}+16q^{28}+9q^{26}+11q^{24}+4q^{22}+10q^{20}+\nn\\
+14q^{18}+33q^{16}+39q^{14}+39q^{12}+24q^{10}+15q^{8}+4q^{6}-2q^{2}-1\nn\\
\alpha^{(3)}_{2,3}=2q^{40}+10q^{38}+30q^{36}+63q^{34}+103q^{32}+139q^{30}+167q^{28}+187q^{26}+208q^{24}+\nn\\
+238q^{22}+283q^{20}+332q^{18}+361q^{16}+344q^{14}+279q^{12}+187q^{10}+100q^{8}+41q^{6}+10q^{4}-1\nn\\
\alpha^{(3)}_{2,4}=-q^{34}-6q^{32}-20q^{30}-47q^{28}-86q^{26}-133q^{24}-187q^{22}-251q^{20}-319q^{18}-378q^{16}-\nn\\
-410q^{14}-399q^{12}-340q^{10}-250q^{8}-154q^{6}-74q^{4}-24q^{2}-4\nn
\\
\alpha^{(3)}_{3,1}=q^{36}+q^{34}+q^{32}-q^{30}-6q^{28}-11q^{26}-20q^{24}-18q^{22}-25q^{20}-23q^{18}-32q^{16}-\nn\\
-25q^{14}-28q^{12}-23q^{10}-25q^{8}-15q^{6}-7q^{4}-q^{2}+1\nn\\
\alpha^{(3)}_{3,2}=q^{46}+6q^{44}+17q^{42}+31q^{40}+38q^{38}+18q^{36}-51q^{34}-181q^{32}-348q^{30}-508q^{28}-\nn\\
-625q^{26}-691q^{24}-700q^{22}-660q^{20}-586q^{18}-495q^{16}-387q^{14}-268q^{12}-154q^{10}-70q^{8}-26q^{6}-10q^{4}-4q^{2}-1\nn\\
\alpha^{(3)}_{3,3}=-2q^{46}-12q^{44}-37q^{42}-77q^{40}-116q^{38}-115q^{36}-22q^{34}+198q^{32}+529q^{30}
+900q^{28}+1212q^{26}+\nn\\
+1391q^{24}+1419q^{22}+1326q^{20}+1153q^{18}+931q^{16}+685q^{14}+449q^{12}+256q^{10}+125q^{8}+53q^{6}+20q^{4}+6q^{2}+1\nn
\ee
\be
\alpha^{(3)}_{3,4}=q^{18}+q^{16}-2q^{12}-2q^{10}-2q^{8}-q^{6}-q^{4}-q^{2}-1\nn
\\
\alpha^{(3)}_{4,1}=q^{28}+6q^{26}+7q^{24}+13q^{22}+8q^{20}+17q^{18}+15q^{16}+21q^{14}+13q^{12}+11q^{10}+9q^{8}+7q^{6}+2q^{4}-q^{2}-1\nn\\
\alpha^{(3)}_{4,2}=-q^{32}+11q^{28}+38q^{26}+82q^{24}+138q^{22}+203q^{20}+267q^{18}+308q^{16}+308q^{14}+\nn\\
+266q^{12}+199q^{10}+130q^{8}+71q^{6}+30q^{4}+8q^{2}+1\nn\\
\alpha^{(3)}_{4,3}=q^{26}-3q^{22}-9q^{20}-16q^{18}-28q^{16}-42q^{14}-52q^{12}-51q^{10}-38q^{8}-18q^{6}-3q^{4}+2q^{2}+1\nn\\
\alpha^{(3)}_{5,1}=q^{28}+4q^{26}+8q^{24}+13q^{22}+13q^{20}+15q^{18}+15q^{16}+17q^{14}+14q^{12}+11q^{10}+10q^{8}+7q^{6}+2q^{4}-q^{2}-1\nn\\
\alpha^{(3)}_{5,2}=-q^{30}-8q^{28}-29q^{26}-74q^{24}-143q^{22}-221q^{20}-281q^{18}-304q^{16}-289q^{14}-\nn\\
-248q^{12}-193q^{10}-135q^{8}-81q^{6}-38q^{4}-12q^{2}-2;\nn\\
\alpha^{(4)}_{0,1}=2q^{26}+5q^{24}+9q^{22}+8q^{20}+2q^{18}-7q^{16}-14q^{14}-14q^{12}-9q^{10}-5q^{8}+3q^{4}+3q^{2}+1\nn\\
\alpha^{(4)}_{0,2}=q^{24}+4q^{22}+10q^{20}+14q^{18}+10q^{16}-3q^{14}-16q^{12}-21q^{10}-15q^{8}-4q^{6}+3q^{4}+3q^{2}+1\nn\\
\alpha^{(4)}_{1,1}=q^{24}+2q^{22}+3q^{20}+3q^{18}+4q^{16}+5q^{14}+3q^{12}+q^{10}-q^{8}-q^{6}-2q^{4}-q^{2}-1\nn\\
\alpha^{(4)}_{1,2}=2q^{36}+7q^{34}+14q^{32}+15q^{30}+10q^{28}+q^{26}-3q^{24}-2q^{22}+8q^{20}+24q^{18}+\nn\\
+40q^{16}+46q^{14}+42q^{12}+30q^{10}+18q^{8}+7q^{6}-2q^{2}-1\nn\\
\alpha^{(4)}_{1,3}=-q^{36}-6q^{34}-19q^{32}-38q^{30}-54q^{28}-59q^{26}-61q^{24}-80q^{22}-132q^{20}-209q^{18}-\nn\\
-279q^{16}-309q^{14}-289q^{12}-229q^{10}-154q^{8}-87q^{6}-39q^{4}-12q^{2}-2\nn\\
\alpha^{(4)}_{2,1}=q^{14}+q^{12}+q^{10}-q^{8}-2*q^{6}-2*q^{4}-q^{2}-1\nn\\
\alpha^{(4)}_{3,1}=q^{20}+q^{18}+2q^{16}+2q^{14}+3q^{12}+4q^{10}+2q^{8}+2q^{6}-1\nn\\
\alpha^{(4)}_{3,2}=q^{38}+5q^{36}+12q^{34}+21q^{32}+26q^{30}+26q^{28}+16q^{26}+q^{24}-16q^{22}-22q^{20}-\nn\\
-12q^{18}+12q^{16}+35q^{14}+47q^{12}+46q^{10}+36q^{8}+19q^{6}+5q^{4}-q^{2}-1\nn\\
\alpha^{(4)}_{3,3}=-2q^{40}-12q^{38}-37q^{36}-78q^{34}-122q^{32}-146q^{30}-129q^{28}-73q^{26}-10q^{24}+\nn\\
+11q^{22}-37q^{20}-135q^{18}-234q^{16}-290q^{14}-282q^{12}-223q^{10}-145q^{8}-76q^{6}-30q^{4}-8q^{2}-1\nn\\
\alpha^{(4)}_{3,4}=q^{20}+2q^{18}+3q^{16}+q^{14}-2q^{12}-4q^{10}-q^{8}+3q^{6}+6q^{4}+5q^{2}+2\nn\\
\alpha^{(4)}_{4,1}=q^{38}+2q^{36}+2q^{34}+q^{32}+q^{30}-7q^{26}-15q^{24}-20q^{22}-15q^{20}-12q^{18}-11q^{16}-\nn\\
-14q^{14}-13q^{12}-11q^{10}-10q^{8}-7q^{6}-2q^{4}+q^{2}+1\nn\\
\alpha^{(4)}_{4,2}=-2q^{40}-11q^{38}-29q^{36}-49q^{34}-58q^{32}-44q^{30}+3q^{28}+84q^{26}+181q^{24}+\nn\\
+260q^{22}+299q^{20}+296q^{18}+266q^{16}+231q^{14}+202q^{12}+170q^{10}+128q^{8}+80q^{6}+38q^{4}+12q^{2}+2\nn\\
\alpha^{(4)}_{5,3}=q^{20}+q^{18}+q^{16}-q^{14}-2q^{12}-2q^{10}-q^{8}-q^{6}-q^{4}-2q^{2}-1\nn
\ee
\be
\beta_1=q^{14}+2q^{12}+2q^{10}+2q^{8}+2q^{6}+4q^{4}+2q^{2}+1\nn\\
\beta_2=q^{28}+2q^{22}+16q^{20}+33q^{18}+46q^{16}+49q^{14}+48q^{12}+40q^{10}+23q^{8}+5q^{6}-3q^{4}-3q^{2}-1\nn\\
\beta_3=-q^{28}-4q^{26}-14q^{24}-39q^{22}-91q^{20}-173q^{18}-269q^{16}-344q^{14}-365q^{12}-320q^{10}-229q^{8}-131q^{6}-58q^{4}-18q^{2}-3\nn\\
\beta_4=q^{26}+4q^{24}+7q^{22}+8q^{20}+7q^{18}+8q^{16}+12q^{14}+17q^{12}+18q^{10}+18q^{8}+13q^{6}+10q^{4}+4q^{2}+1\nn\\
\beta_5=-2q^{34}-12q^{32}-36q^{30}-74q^{28}-118q^{26}-158q^{24}-190q^{22}-220q^{20}-246q^{18}-258q^{16}-\nn\\
-244q^{14}-202q^{12}-143q^{10}-86q^{8}-44q^{6}-19q^{4}-6q^{2}-1\nn\\
\beta_6=q^{14}+q^{12}+2q^{10}+2q^{8}+4q^{6}+3q^{4}+2q^{2}+1\nn\\
\beta_7=-2q^{24}-5q^{22}-10q^{20}-13q^{18}-15q^{16}-11q^{14}-3q^{12}+7q^{10}+12q^{8}+13q^{6}+9q^{4}+4q^{2}+1\nn\\
\ee
}

\end{document}